\documentclass[preprint,12pt]{emulateapj}

\usepackage{color}
\usepackage{CJKutf8}
\begin{document}
\title{Physical Structures of the Type Ia Supernova Remnant N103B}
%
%

\author{Chuan-Jui Li \begin{CJK}{UTF8}{bsmi}(李傳睿)\end{CJK}\altaffilmark{1,2}, You-Hua Chu \begin{CJK}{UTF8}{bsmi}(朱有花)\end{CJK}\altaffilmark{1,3}, Robert A.\ 
Gruendl\altaffilmark{3}, Dan Weisz\altaffilmark{4}, Kuo-Chuan Pan \begin{CJK}{UTF8}{bsmi}(潘國全)\end{CJK}\altaffilmark{5,6}, \\ Sean D. Points\altaffilmark{7}, Paul M.\ Ricker\altaffilmark{3}, R.\ Chris Smith\altaffilmark{7}, Frederick M.\ Walter\altaffilmark{8} }

\affil{$^1$ Institute of Astronomy and Astrophysics, Academia Sinica, Taipei, Taiwan, R.O.C.\\
$^2$ Graduate Institute of Astrophysics, National Taiwan University, Taipei, Taiwan, R.O.C.\ \\
$^3$ Department of Astronomy, University of Illinois at Urbana-Champaign, Urbana, U.S.A. \\
$^4$ Department of Astronomy, University of California, Berkeley, Berkeley, U.S.A. \\
$^5$ Departement Physik, Universit\"{a}t Basel, Basel, Switzerland \\
$^6$ Department of Physics and Astronomy, Michigan State University, East Lansing, U.S.A.\\
$^7$ Cerro Tololo Inter-American Observatory, La Serena, Chile \\
$^8$ Department of Physics and Astronomy, Stony Brook University, New York, U.S.A.}

%
%

%
%
%
\begin{abstract}
N103B is a Type Ia supernova remnant (SNR) projected in the outskirt of the superbubble
around the rich cluster NGC\,1850 in the Large Magellanic Cloud (LMC).  We have obtained
H$\alpha$ and continuum images of N103B with the \emph{Hubble Space Telescope} 
(\emph{HST}) and high-dispersion spectra with 4m and 1.5m telescopes at Cerro Tololo 
Inter-American Observatory.  The \emph{HST} H$\alpha$ image exhibits a complex system 
of nebular knots inside an incomplete filamentary elliptical shell that opens to the east where 
X-ray and radio emission extends further out.  Electron densities of the nebular knots, 
determined from the [S II] doublet, reach 5300 cm$^{-3}$, indicating an origin of circumstellar medium, rather than interstellar medium. The high-dispersion spectra reveal three kinematic components in N103B: (1) a narrow component with [N II]6583/H$\alpha$ $\sim$ 0.14 from 
the ionized interstellar gas associated with the superbubble of NGC\,1850 in the background, 
(2) a broader H$\alpha$ component with no [N II] counterpart from the SNR's collisionless 
shocks into a mostly neutral ambient medium, and (3) a broad component, $\Delta V$ 
$\sim$ 500 km s$^{-1}$, in both H$\alpha$ and [N II] lines from shocked material in the 
nebular knots.  The Balmer-dominated filaments can be fitted by an ellipse, and we adopt 
its center as the site of SN explosion.  We find that the star closest to this explosion center 
has colors and luminosity consistent with a 1 $M_\odot$ surviving subgiant companion as 
modelled by Podsiadlowski.  Follow-up spectroscopic observations are needed to confirm 
this star as the SN's surviving companion.
\end{abstract}
\subjectheadings{circumstellar matter --- ISM: supernova remnants --- Magellanic Clouds
}
\maketitle
%
%
%
%
\section{Introduction}  \label{sec:intro}

While Type Ia supernovae (SNe) are widely known as standardizable candle for cosmological 
distances, the nature of their progenitor systems is still in debate.  Owing to the absence of 
strong hydrogen lines in their spectra, it has been suggested that their progenitors should be 
hydrogen-deficient stars.  The most plausible candidates for the progenitors are the compact 
and degenerate stellar remnants of low mass stars known as white dwarfs (WDs).  It is 
believed that Type Ia is the only SN type that results from the thermonuclear disruption of a 
carbon-oxygen WD.  Two main scenarios have been suggested: a single-degenerate (SD) 
origin in which a WD has accreted matter from a non-degenerate binary companion until its
mass reaches the Chandrasekhar limit of $\sim$ 1.4 {\it M$_\sun$} 
\citep{whelan1973, nomoto1982},
and a double-degenerate (DD) origin that requires the merger of two WDs 
\citep{iben1984, webbink1984}. 

\begin{figure*}  
\epsscale{1}
\hspace*{-1cm}\plottwo{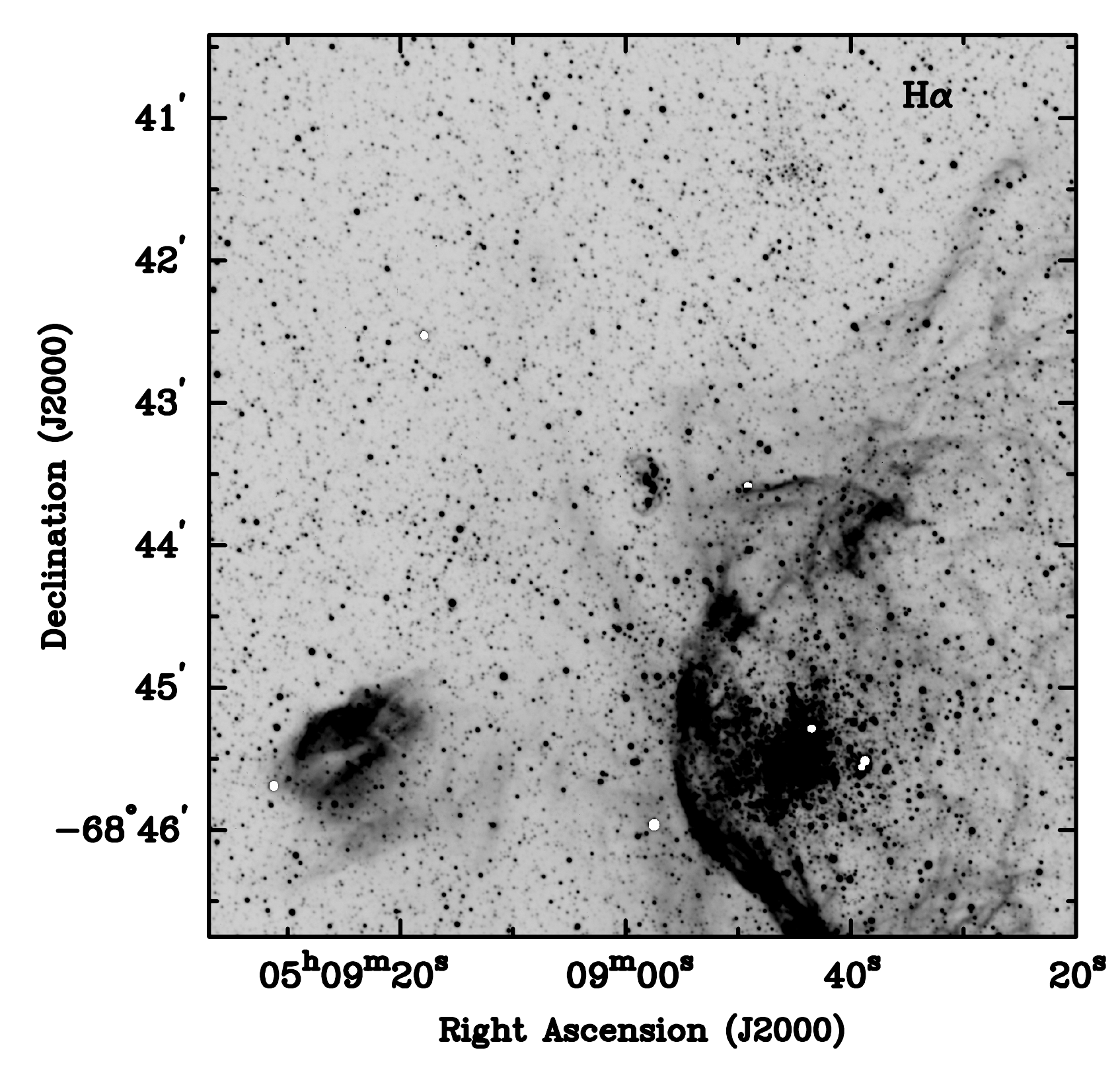}{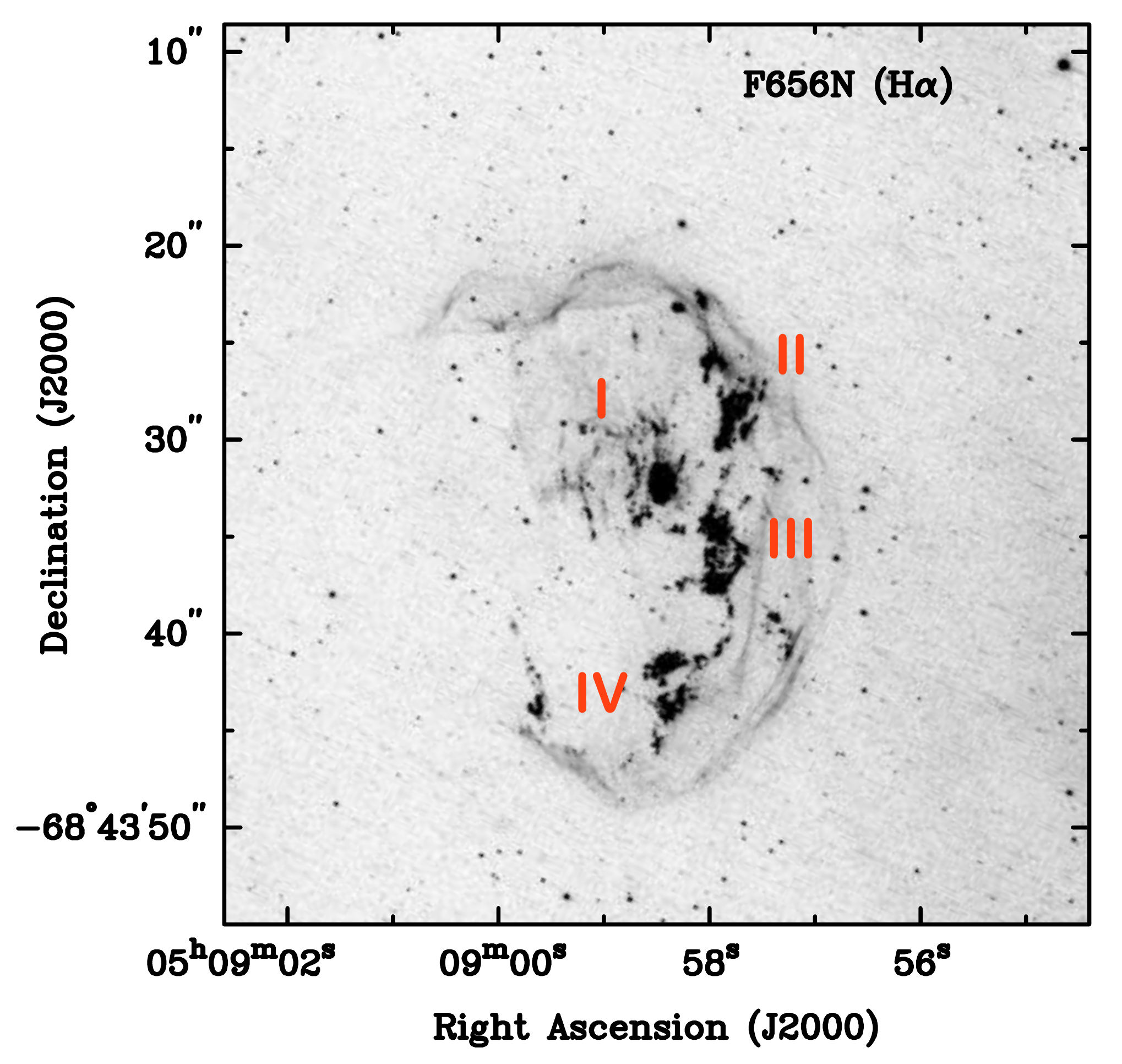}
\caption{ (Left) H$\alpha$ image of N103B obtained with the CTIO Blanco 4m telescope 
and the MOSAIC\,II camera showing the large-scale environment of the SNR.   (Right) 
A high-resolution H$\alpha$ image of N103B obtained with the \emph{HST} WFC3. 
Four prominent groups of knots, I--IV, are identified and marked, each consisting of 
multiple concentrations of knots.  There are strings of knots connecting these four 
groups of knots.  No nebular knots are seen in the southeast quadrant of the SNR.}
\label{fig:ground_hst}
\end{figure*}

In the SD scenario, the non-degenerate companion may be a main-sequence (MS) star, 
a red giant (RG), or a helium star.  The close binary interactions between the 
non-degenerate companion and the WD may result in mass loss and produce a dense 
circumstellar medium (CSM).  The non-degenerate companion can survive the SN explosion and
may even be bright enough to be detected \citep{marietta2000, pan2014}. 
In the DD scenario, no dense CSM  or detectable surviving companion is 
expected.
Therefore, if a surviving companion or a dense CSM is detected, the 
SD origin of a Type Ia SN can be affirmed.  While no surviving companions of Type Ia 
progenitors have been unambiguously identified, a dense CSM has
been detected in the Type Ia Kepler SN remnant (SNR) and used to suggest a SD origin 
 \citep{vandenbergh1973, vandenbergh1977, dennefeld1982, 
blair1991,williams2012}. 

The SNR 0509-68.7, also known as N103B, in the Large Magellanic Cloud (LMC) has
been suggested to be an old cousin of the Kepler SNR \citep{williams2014}.
The SNR nature of N103B was first identified by its nonthermal radio emission and 
later confirmed by its diffuse X-ray emission \citep{mathewson1983}.  
As N103B is projected near the ionized superbubble around the cluster NGC\,1850, 
it has been suggested to be associated with a core-collapse SN 
\citep{chu1988}.
However, the X-ray spectra of N103B show SN ejecta abundances consistent with 
those of Type Ia SNe \citep{hughes1995}; furthermore, optical spectra of 
the light echo of N103B's SN exhibit Type Ia spectral characteristics, i.e., lacking 
hydrogen lines \citep{rest2005, Restprep2016}.
It is now well recognized that N103B is a Type Ia SNR.

N103B has been included in our study of Type Ia SNRs in the LMC.  We have obtained
\emph{Hubble Space Telescope} (\emph{HST}) H$\alpha$ (F656N) images to study 
the physical structure of the SNR, as well as $BVI$ (F475W, F555W, and F814W) 
images to search for surviving companions of the SN progenitors and to study the 
underlying stellar population.  We have also obtained high-dispersion echelle spectra 
of N103B to use its kinematic properties to differentiate between the CSM and
interstellar medium (ISM) and to diagnose SNR shocks.  This paper reports our 
analysis of the physical structure of the SNR N103B and 
search for a surviving companion of its SN progenitor. 

\begin{figure*}  
\epsscale{0.6}
\plotone{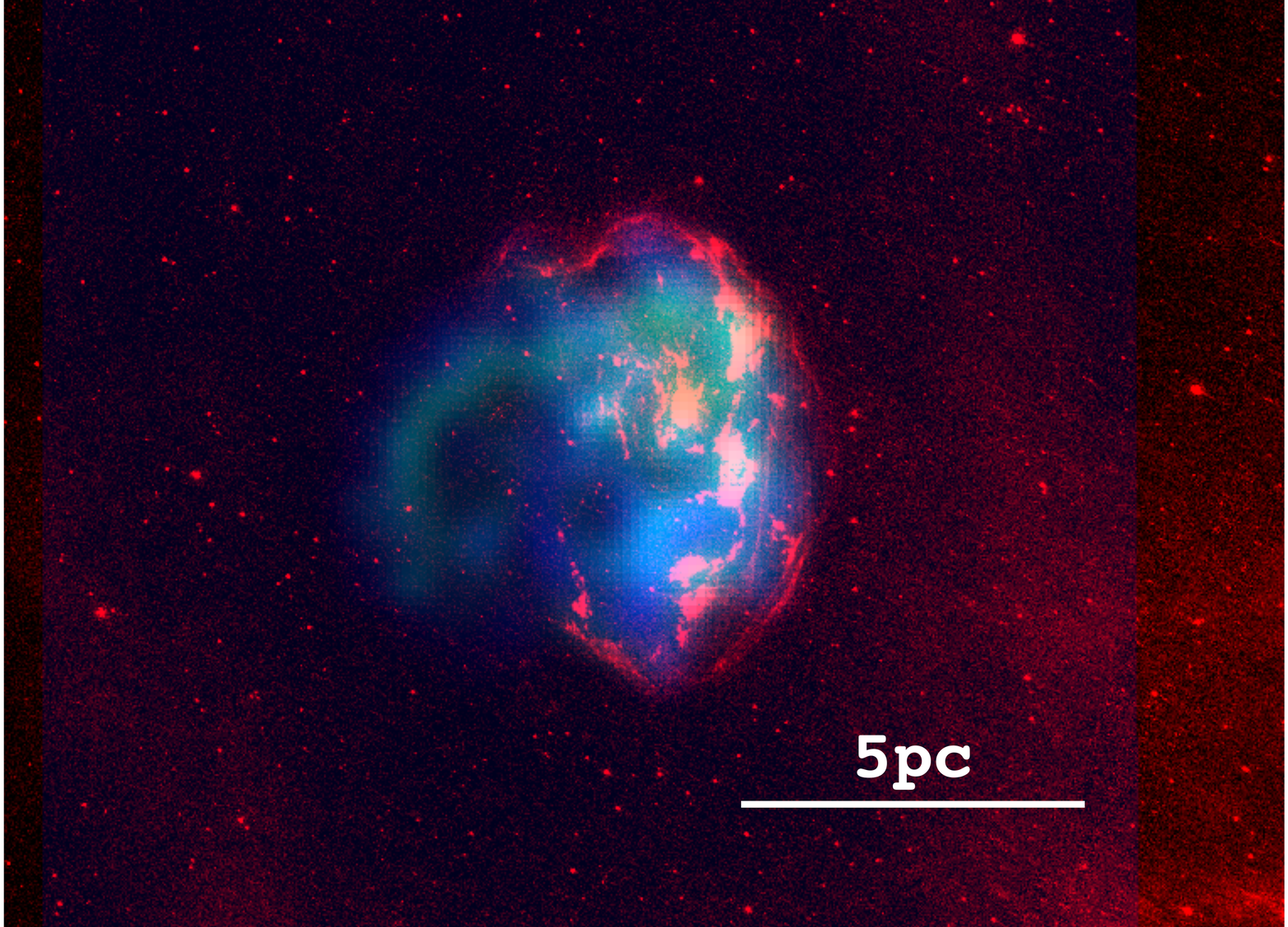}
\caption{
Color composite image of N103B with the \emph{Chandra}  ACIS images of
the 0.9-10 keV band in blue and 0.6-0.9 keV band  in green, and 
\emph{HST} WFC3 H$\alpha$ image in red. The X-ray emission extends
further toward the east by $\sim$15\arcsec\ (or 3.75 pc) where no H$\alpha$ counterpart
is detected.  The diffused H$\alpha$ background 
is stronger to the southwest of the SNR, toward the superbubble of NGC\,1850.
(The X-ray images are from The Chandra Catalog of Magellanic Cloud SNR,
http://hea-www.harvard.edu/ChandraSNR/snrcat$_{-}$lmc.html)}
\label{fig:x_ray}
\end{figure*}

The paper is organized as follows.  Section 2 describes observations used in 
this study, Section 3 examines the environment and morphology of N103B, 
Section 4 presents a quantitative analysis of surface brightnesses and 
densities of nebular knots in the SNR interior, and Section 5 uses
detailed kinematic properties to differentiate between the SNR and background
gas components.  Section 6 discusses the structure and formation of the N103B
SNR, a possible candidate for surviving companion, and a comparison with 
the Kepler SNR.  Finally, a summary is given in Section 7.

\section{Observations} \label{sec:obs}  
\subsection{\emph{HST} Observations}

The \emph{HST} H$\alpha$ images of N103B were obtained in Program 13282 (PI: Chu) on 2013 July 11,
using the UVIS channel of Wide Field Camera 3 (WFC3) and the F656N filter.  The 162$''$$\times$162$''$
field of view is large enough to cover not only the entire N103B SNR but also part of the nearby superbubble
around the NGC1850 cluster.  The pixel size, 0\farcs04, corresponds to 0.01 pc in the LMC (at 50 kpc distance).
 
The H$\alpha$ imaging observations were dithered with the WFC3-UVIS-GAP-LINE pattern for 3 points and
point spacings of 2\farcs414.  The total exposure time for each position in N103B was 1350 s.
All exposures used the FLASH=11 option to make sure a background of roughly 10-12 e$^-$ was present to 
compensate for charge transfer efficiency (CTE) issues in WFC3.

Using the \emph{HST} we have also obtained F475W $(B)$, F555W $(V)$ and 
F814W $(I)$ band images of N103B to study it's underlying stellar population. 
The UVIS channel of WFC3 was used for continuum-band images. 
All continuum-band observations have a 162$''$$\times$162$''$ field of view as F656N 
observation. The F475W and F814W observations each had
a total exposure time of 1050 s. 
The F555W observation had a total exposure time of 1117 s. The FLASH=5 option 
was used for the F475W observation, and the FLASH=4 option was used for the 
F814W observation to mitigate the CTE issues in WFC3. 

We used \texttt{DOLPHOT}, a version of \texttt{HSTPHOT}
with \emph{HST}-specific modules \citep{dolphin2000},
to carry out point-spread-function photometry on each of 
the \emph{HST} images.  The photometric parameters
were selected following the recommendations of \citet{williamsbf2014}. 
The catalog of detected objects was filtered to include only well-measured stars using the criteria of signal-to-noise ratio $>$ 5, $sharp^2 < 0.1$, and
$crowd < 1.0$, as recommended by \citet{williamsbf2014}.  
Definitions of these parameters can be found in \citet{dolphin2000}.
These photometric measurements were used to construct 
color-magnitude diagrams (CMDs) in order to study the stellar population.

\begin{figure*}  
\epsscale{0.8}
\centering
\hspace*{-0.5cm}\plotone{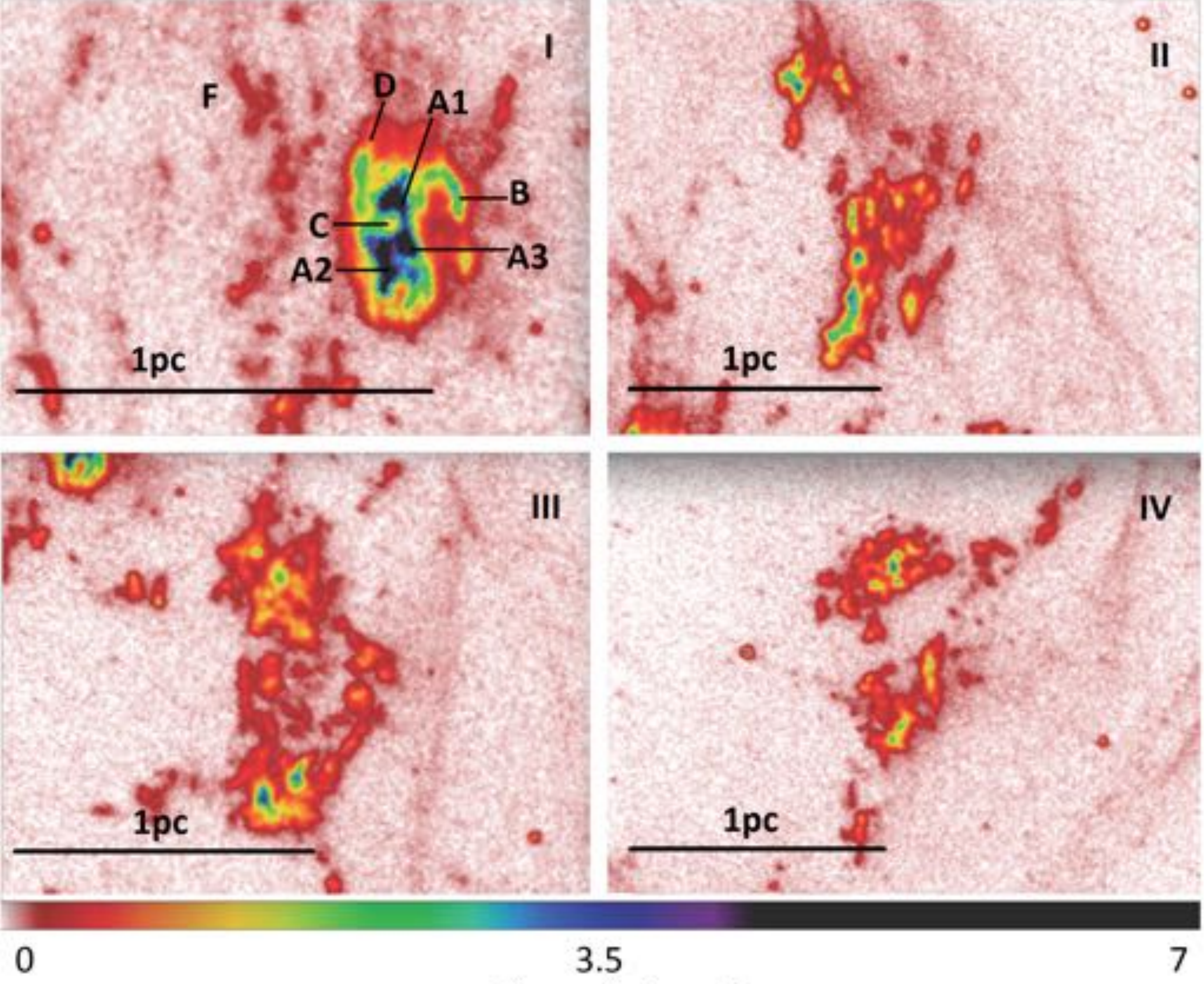}
\caption{Close-up H$\alpha$ images of the four groups of nebular knots.  
The intensity-color scale bar is displayed at the bottom in units of 10$^{-13}$ erg\,cm$^{-2}$ s$^{-1}$ arcsec$^{-2}$.}
\label{fig:close_up}
\end{figure*}

\subsection{CTIO Observations}
\subsubsection{Imaging}

The Cerro Tololo Inter-American Observatory (CTIO) Blanco 4m 
H$\alpha$ image of N103B was taken from
the Magellanic Cloud Emission-Line Survey 2 (MCELS2), which 
used the MOSAIC\,II camera and covered the entire LMC.  
The MOSAIC\,II camera consists of eight SITe 4096$\times$2048 
CCDs with a pixel size of 0\farcs27 pixel$^{-1}$ and a total 
field-of-view $\sim$ 36$'$$\times$36$'$.  For MCELS2, each field was 
imaged with 1 and 10 s short exposures and  3$\times$300 s 
long exposures dithered to compensate the gaps between CCDs.
The SuperMACHO pipeline software was used for bias subtraction, flat 
fielding, and distortion correction.  The MCELS2 survey was made
in the H$\alpha$ line only, as opposed to multiple lines and continuum
bands for MCELS1 (Smith and MCELS team 1999).

\subsubsection{Long-Slit Echelle Spectroscopic Observations}

High-dispersion, long-slit spectra of N103B were obtained with the echelle spectrograph on the
CTIO Blanco 4m telescope on 1995 January 19. A 79 line mm$^{-1}$ echelle grating was used.
A post-slit H$\alpha$  filter ($\lambda_c=$ 6563 \AA \ and $\Delta \lambda=$ 75 \AA)
was inserted and the cross-disperser was replaced by a flat mirror to isolate a single order 
and to allow a maximum ($\sim$4$'$) slit coverage.  The H$\alpha$ filter was wide enough 
to transmit both the H$\alpha$ and the neighboring [\ion{N}{2}] $\lambda\lambda$6548, 6583 lines.  
The long-focus red camera and a Tek 2048 $\times$2048 CCD were used to record data.
The pixel size was 24 $\mu$m, corresponding to 0.082 \AA \,pixel$^{-1}$ 
(3.75 km s$^{-1}$ pixel$^{-1}$) along the direction of dispersion and 0\farcs267 pixel$^{-1}$ 
perpendicular to the dispersion.   A 250 $\mu$m (1\farcs65) slit width was used, resulting
in an instrumental FWHM of about 14 km s$^{-1}$.  

Two slit positions were observed, along position angles 0\degr\ and 44\degr, respectively.  
The slit was long enough to cover not only the entire SNR N103B but also part of
the nearby superbubble around the NGC1850 cluster.  The journal of echelle observations
is given in Table 1.

\begin{deluxetable}{ccccc}
\tablecolumns{2}
\tabletypesize{\scriptsize}
\tablewidth{0pc}
\tablecaption{CTIO 4m Echelle Observations of N103B}
\tablehead{
 RA & Dec & Date & Position Angle& Exposure\\  
 &&&(degree)  & (s)
}
\startdata 
5:09:11.96&-68:47:14.60&1995 Jan 19 &  44 &  ~600\\
5:09:12.26&-68:47:14.60&1995 Jan 19 &  44 & ~600\\
5:08:58.94&-68:43:36.20&1995 Jan 19 &  ~0 & ~300\\
5:08:59.05&-68:43:38.80&1995 Jan 19 &  ~0& 1200\\
5:08:58.93&-68:43:37.50&1995 Jan 19 &  ~0& 1200
\enddata
\tablecomments{Position angles were measured counterclockwise from the north.  }
\end{deluxetable}

 The data were bias subtracted and flat-field corrected at the telescope, and later processed
 using the IRAF software for cosmic ray rejection, geometric distortion correction, and 
 wavelength calibration.   The solutions for geometric distortion and wavelength calibration 
 were derived from Th-Ar lamp observations.  Finally, the telluric H$\alpha$ emission in each
 frame was used as reference to fine-tune the absolute wavelength calibration.  Heliocentric 
 velocities are used in this paper.

\subsubsection{Multi-Order Echelle Spectroscopic Observations}
High-dispersion multi-order echelle observations of N103B were obtained with the 
CHIRON spectrometer \citep{tokovinin2013} on the SMARTS 1.5 m telescope at 
CTIO on 2015 February 11.  
Chiron is fed by a fiber with a 2.7 arcsec diameter on the sky.
Only the brightest group of nebular knots in N103B was observed.  Three 15 min 
exposures were made with the fiber centered on these bright nebular knots.
The sky background was observed for 15 min with the fiber centered
at about 11 arcsec west of the nebular knots.
 
The multi-order spectrum of the nebular knot, extracted with sky 
subtraction, covers a wavelength range of 4500 -- 8900 \AA.  
Nebular emission lines detected include H$\alpha$, [\ion{N}{2}]$\lambda$6583, 
[\ion{S}{2}]$\lambda\lambda$6716, 6731, [\ion{O}{1}]$\lambda\lambda$6300, 6364, 
[\ion{O}{2}]$\lambda$7318, and [\ion{O}{3}]$\lambda$$\lambda$4595, 5007.
The H$\beta$ line is detected with a low S/N.


\section{Environment and Morphology of N103B}

N103B is superposed on diffuse H$\alpha$ emission in the outskirt of the superbubble 
around NGC\,1850, as shown in the  left panel of Figure \ref{fig:ground_hst}.  In the H$\alpha$-emitting
nebula catalog of  \citet{henize1956}, the bright eastern rim of the superbubble 
was designated as nebula N103B, while the small \ion{H}{2} region to the east 
of the superbubble was designated as nebula N103A. In this naming scheme, ``N103B'' 
would be a misnomer for the SNR, as the SNR is just a small part of the nebula N103B. 
For brevity, we will continue using the name ``N103B'' for the SNR in this paper, although 
its proper name should be SNR B0509-68.7 or J050854-684447.

N103B is projected at $\sim$40 pc from the rich cluster NGC\,1850.  The superbubble
around NGC\,1850 appears to show a breakout structure at the rim closest to N103B.
Exterior to the superbubble, H$\alpha$ emission shows a complex structure with various
streamer-like features superposed on a diffuse component.  N103B is projected between
two streamer-like features.  As N103B is closer to the western feature, the diffuse H$\alpha$
background is the strongest to the southwest and west of the SNR, and fades away on 
the east side of the SNR.

In deep ground-based H$\alpha$ images, the SNR N103B appears like the letter ``C'' 
opening  toward the east where no detectable H$\alpha$ emission exists but radio 
and X-ray emissions extend further by $\sim$15\arcsec\ \citep{lewis2003} 
, as seen in Figure \ref{fig:x_ray}.
The \emph{HST} H$\alpha$ image resolves N103B into an elliptical shell delineated 
by fine filaments with an opening to the east and prominent groups of nebular knots in 
the shell interior, as shown in the right panel of Figure \ref{fig:ground_hst}.

 
Four major groups of nebular knots exist in the shell interior, all located in the western 
half of the shell.  
Among these four, Group I is the most prominent and the closest to the
shell center, only offset by $\sim$3$''$ NW.  Its complex internal structure consists 
of both knots and arcs.  It appears to be connected with strings of fainter knots and 
structured diffuse emission that extend to the eastern side of the shell center.
Groups II--IV, located to the west of Group I, all show multiple concentrations of knots;
furthermore, they appear to be connected with a curved string of knots.
Close-up H$\alpha$ images of these four group of knots are shown in Figure \ref{fig:close_up}.
In addition to these four major groups of knots, a network of faint filaments extends from
the shell center to 6$''$ north, and a small number of knots are projected near the 
filaments along the shell rim.  No prominent nebular knots are seen toward
the eastern part of the remnant, except a small U-shaped feature near the southeast
rim of the shell.  It is particularly curious that the interior of the southeast quadrant
appear to be totally devoid of nebular knots.

\section{Surface Brightness and Densities of Nebular Knots}

The \emph{HST} H$\alpha$ image can be used to determine surface brightness and 
rms density of nebular features, if their emitting path lengths can be reasonably assumed.
We use the complex features in Group I to illustrate our analyses.  The overall 
structure of Group I may be described as small dense knots embedded in a broader diffuse 
gas component (see Figure \ref{fig:close_up}).  
We select the three brightest knots, designated as A1--A3, and assume 
that their emitting path lengths are similar to their diameters.  The northwest arc is 
designated as B, and the width and length of the arc have been adopted as the emitting 
path lengths to assess a possible range of electron densities.  The central region not 
covered by bright knots is still very bright, and we designate it as C and adopt the spatial 
extent of its isophote  as the emitting path length.  The diffuse knots in the immediate outskirts 
of Group I are designated as D, and their emitting path lengths are assumed to be the
same as the diameters of the knots.  The faint diffuse filaments to the east of Group I 
are designated as F, and the width and length of each filament are adopted as emitting
path lengths to derive the upper and lower limits of its rms density.

\begin{deluxetable*}{cccccc}
\tabletypesize{\scriptsize}
\tablewidth{0pc}
\tablecaption{Physical Structures of N103B from Emission Measure}
\tablehead{ ID  & Position & Surface Brightness & n$_e$ & Mass \\
 & & (10$^{-13}$ erg\,cm$^{-2}$ s$^{-1}$ arcsec$^{-2}$) & (cm$^{-3})$ & (10$^{-3}${\it M$_\sun$}) 
}
\startdata
 ISM1& 30$''$SE & 0.02 &  &   \\
 ISM2 & 30$''$NE & 0.02 &  &   \\
 ISM3 & 30$''$NW & 0.03 & &   \\
 ISM4 & 30$''$SW & 0.04 &  &   \\ \\
Group I &  &  &  &  & \\
 & A1 &  6.9 &   2250 $\pm$ 200  & 2.2 $\pm$ 0.2  \\
 & A2 &  6.5 &   2000 $\pm$ 600  & 3.9 $\pm$ 1.2  \\
 & A3 &  6.5 &   2150 $\pm$ 250  & 2.2 $\pm$ 0.3  \\\\
 & B  & 1.3--2 & 600--1000  & 240\\
 & C  & 4.6 &  750--1000 & \\
 & D  & 0.3 & 300--400 & \\\\
 & F  &  0.1--0.5 & 250--500 & 100\\\\
Group II  &  &  &  &  & \\
 & Knots & 0.6--1.7 &  600--900 & 200 \\\\
Group III  &  &  &  &  & \\
 & Knots & 0.4--1.5 & 500--1000 &    240\\\\
Group IV  &  &  &  &  & \\
& Knots & 0.5--1.5 &  600--900& 130 
\enddata
\tablenotetext{*}{The morphological features are referred to Figure \ref{fig:close_up}. 
The total mass of nebular knots in N103B is about 1 {\it M$_\sun$}.}
\end{deluxetable*}

The H$\alpha$ surface brightness (SB) of a 10$^4$ K ionized gas region 
can be expressed as 
\begin{equation}
\mathrm{SB} \sim 1.9\times10^{-18}n_\mathrm{e}^2 L_\mathrm{pc} 
\mathrm{\,\,\,\,erg\,cm}^{-2} \mathrm{s}^{-1} \mathrm{arcsec}^{-2},
\end{equation}
where $n_\mathrm{e}$ is rms electron density and $L_\mathrm{pc}$ is the emitting 
path length along the line of sight. 
We assume that the structure of Group I can be described as dense knots in a diffuse
gas region; thus, the emission from the knots and
arc features is superposed on a ``local background" component emitted by the diffuse gas.
We have measured the background-subtracted surface brightnesses of the
knots and arc features, converted them to emission measures (EM 
$\equiv$ $n_\mathrm{e}^2 L_\mathrm{pc}$), and listed them in Table 2.   
The C feature is between knots, so
it represents emission from the diffuse gas component, and the surface brightness 
of the regions surrounding Group I is used as background and subtracted from that
of the C feature.
Note that the global background due to the ionized gas associated with the superbubble
of NGC\,1850 has been measured in four locations at 30$''$ from the center of the
H$\alpha$ shell, denoted as ISM1-4 in Table 2; this global background has been
interpolated and subtracted from all measurements.
The rms densities of the features, derived using the emitting path lengths described 
above, are also given in Table 2.

Inside the brightest concentration of knots (Group I), the brightest emission 
comes from the A1--A3 knots with surface brightness up to  
6.9$\times$10$^{-13}$ erg\,cm$^{-2}$ s$^{-1}$ arcsec$^{-2}$.
The densities derived for these knots are high, $n_{\rm e} \sim$ 2000 $\pm$ 600 cm$^{-3}$.
The diffuse component near the center of Group I, feature C, has $n_{\rm e} \sim$
 750--1000 cm$^{-3}$.  The knots outside the main body of Group I,
feature D, has $n_{\rm e}$ 300--400 cm$^{-3}$.
The knots in Groups II--IV all have $n_{\rm e}$ in the range of 500--1000 cm$^{-3}$.  
The faint filaments to the east of Group I have the lowest $n_{\rm e}$, 
$\sim$250--500 cm$^{-3}$.
Such high densities and nebular morphology are more consistent with a
CSM than an ISM (see Table 1.3 of Draine 2011).

\begin{figure*}  
\epsscale{0.8}
\plotone{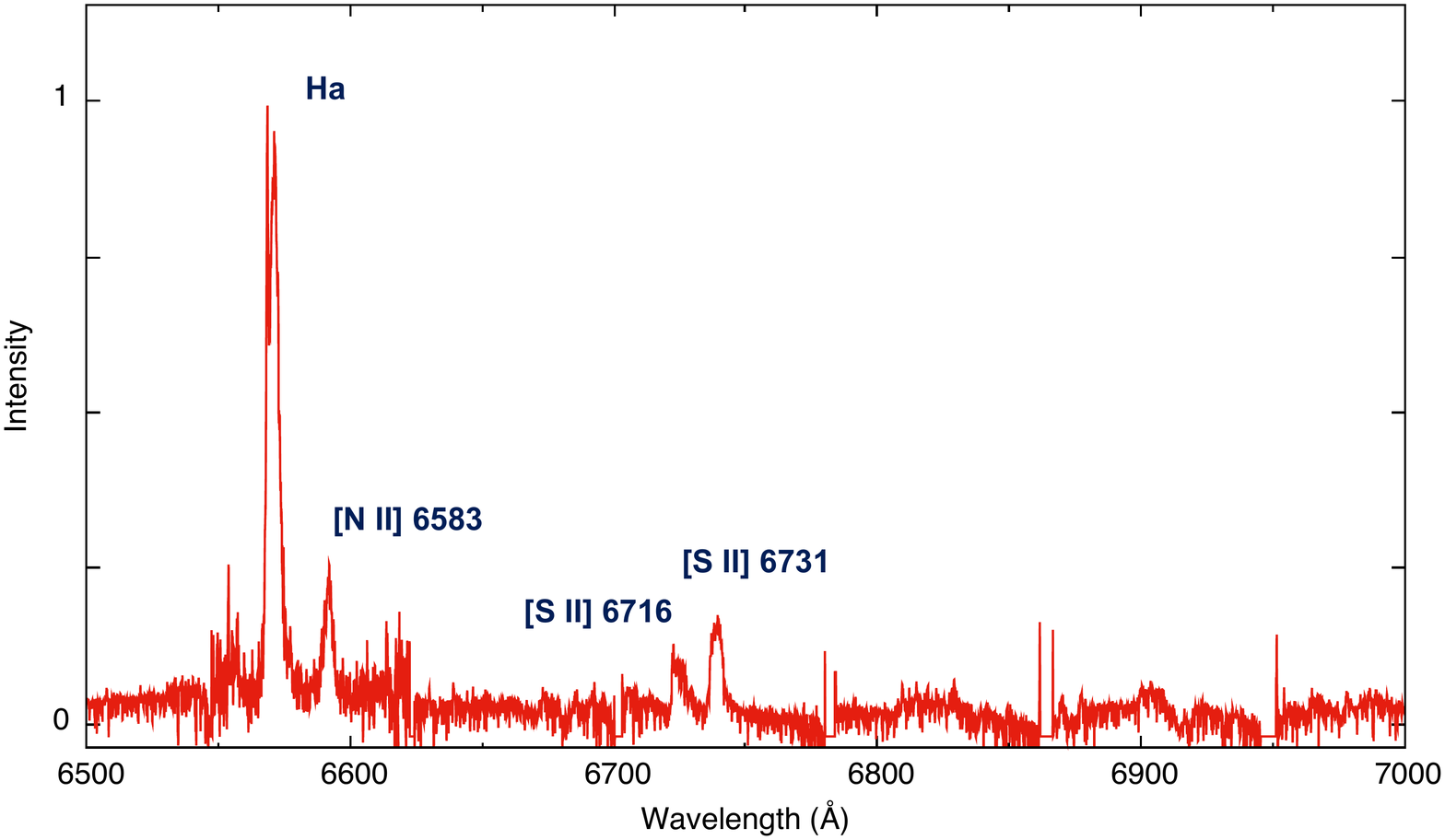}
\caption{High-dispersion multi-order echelle spectrum of N103B obtained 
with the CHIRON spectrometer on the SMARTS 1.5 m telescope at CTIO. The 
H$\alpha$, [N II] $\lambda$6583, and [S II] $\lambda\lambda$6716, 6731 doublet 
are marked. The slit was centered on the brightest nebular knots (Group I) in N103B.
The spectral gaps between orders can be seen at roughly 6465, 6620, 6700, 6780, 
6865, and 6950 \AA.}
\label{fig:chiron}
\end{figure*}

Using the high-dispersion spectra obtained with the CHIRON spectrometer on 
the SMARTS 1.5 m telescope, we have derived an electron density of 
$\sim$5300 cm$^{-3}$ from the [\ion{S}{2}] $\lambda$6716/$\lambda$6731
ratio of the Group I feature (a blend of the knots and the diffuse component) 
as shown in Figure \ref{fig:chiron}.
Electron densities derived from [\ion{S}{2}] doublet are expected to be higher 
than the rms densities, as forbidden line emission is weighted toward high-density
regions.  The rms densities of the knots being lower than that derived from 
the [\ion{S}{2}] doublet indicate that the physical sizes of the knots are actually
smaller than we assumed, i.e., the filling factor is smaller than 1.

We may conclude that the nebular knots in N103B have densities ranging from
$\sim$500 to 5000 H-atom cm$^{-3}$.  The diffuse gas surrounding the knots 
has rms densities ranging from a few hundred to $\sim$1000 H-atom cm$^{-3}$.
Using the rms densities and the approximate sizes of the features, we can
estimate the gas mass in each feature.  These masses are given in the last
column of Table 2.  The total mass of the ionized gas is $\sim$ 1 $M_\odot$.

The recombination timescale for an ionized gas is ($n_\mathrm{e} \alpha_\mathrm{A}$)$^{-1}$
$\sim$ 7.6$\times$10$^{4}$ $n_\mathrm{e}^{-1}$ yr, where $n_\mathrm{e}$ is the
electron density in units of cm$^{-3}$ and $\alpha_\mathrm{A}$ is the recombination
coefficient, $\sim$4.18$\times$10$^{-13}$ cm$^3$ s$^{-1}$ at 10$^4$ K 
\citep{seaton1959, burgess1964, pengelly1964, hummer1987} .
It is interesting to note that the recombination timescales for the nebular knots 
in N103B are as short as 15 to 150 yr, much shorter than the age of the SNR itself,
which is 860 yr derived from the SN light echos \citep{rest2005}.
Such short recombination timescale indicates that the ionization of the knots occurred
only recently, and are most likely caused by the passage of shocks.
It is conceivable that some dense circumstellar material in the form of nebular
knots has already recombined and become invisible.  The total mass in the
CSM can be much higher than the 1 $M_\odot$ estimated above.

\section{Spatiokinematically Differentiating SNR and Background Components} 

\begin{figure}  
\epsscale{3}
\plottwo{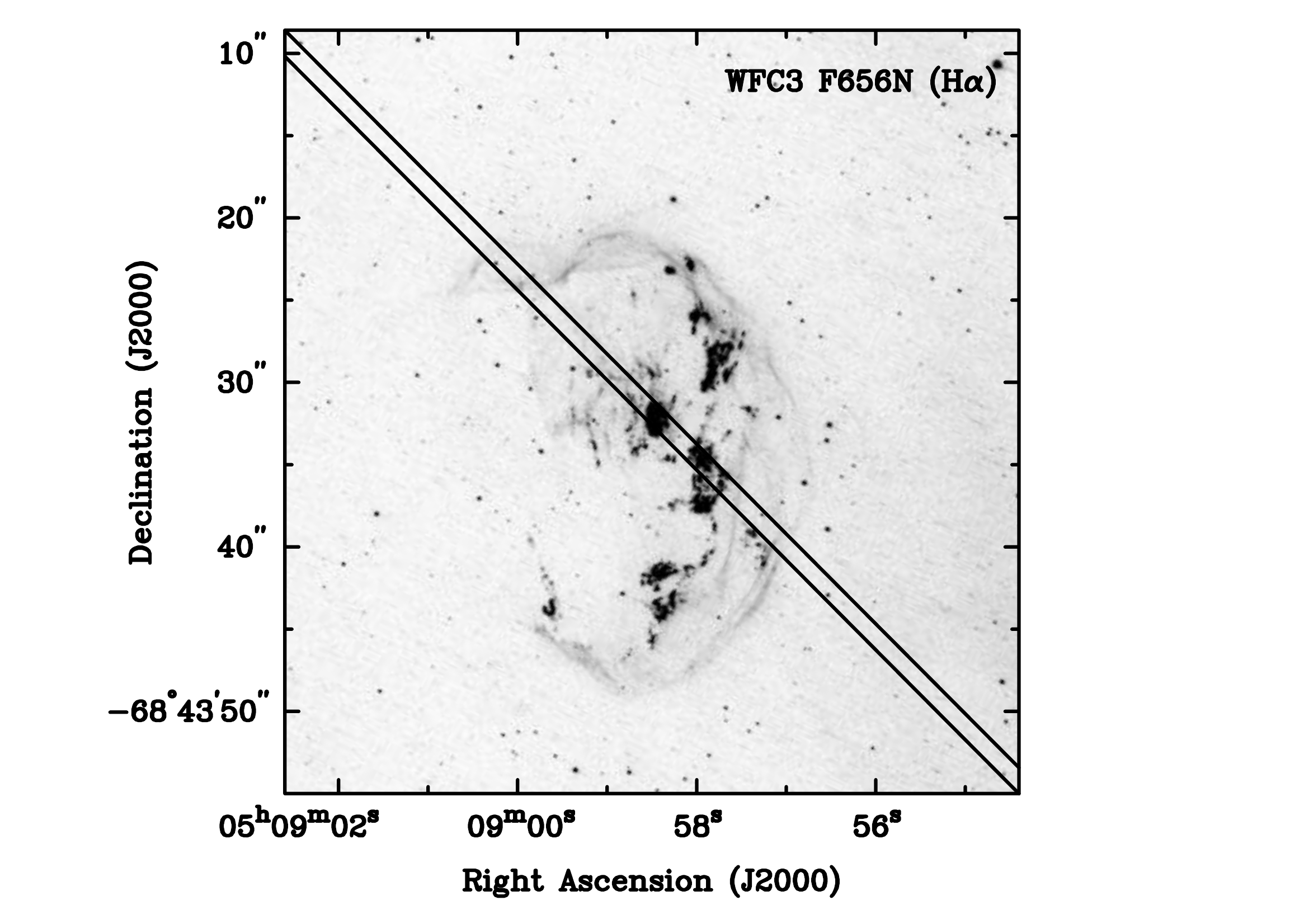}{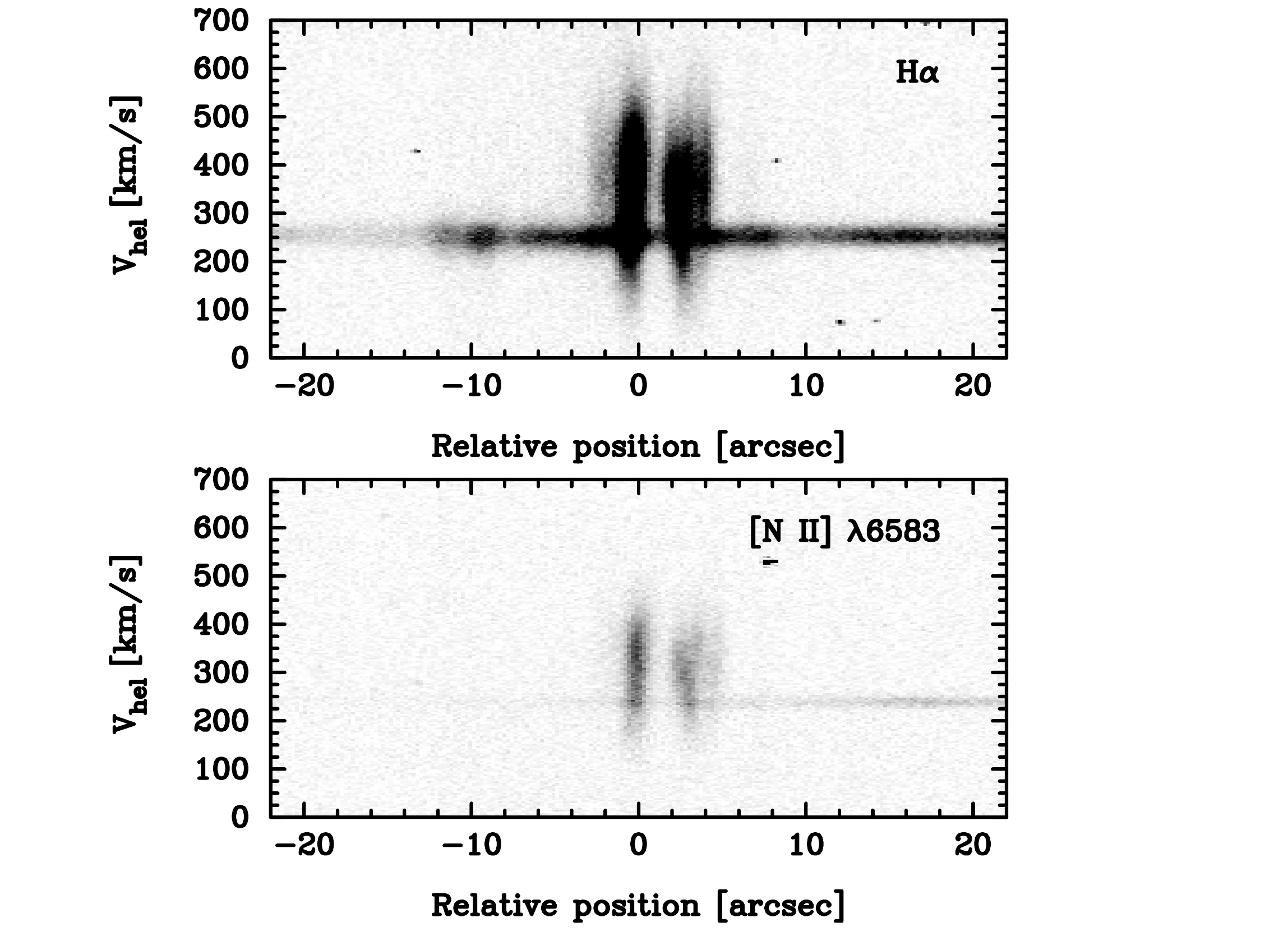}
\caption{(Top) The \emph{HST} WFC3 H$\alpha$ image of N103B marked
with the slit position along position angle 44\degr.
(Bottom) The CTIO long-slit echellogram  of H$\alpha$ and [N II] $\lambda$6583 emission lines. 
Three distinct components are visible: (1) a narrow background component in both H$\alpha$ 
and [\ion{N}{2}] lines extending throughout the entire slit length, (2) a brighter and broader 
H$\alpha$ component without [\ion{N}{2}] counterpart at velocities similar to that of the 
background component at locations corresponding to the H$\alpha$ filamentary shell,
and  (3) a broad component in both H$\alpha$ and [\ion{N}{2}] lines with velocities extending
over $\Delta V$ up to $\sim$ 500 km s$^{-1}$ at locations corresponding to the dense nebular knots.}
\label{fig:slit_WE} 
\end{figure}

\begin{figure*} 
\epsscale{0.67}
\hspace*{-1cm}\plotone{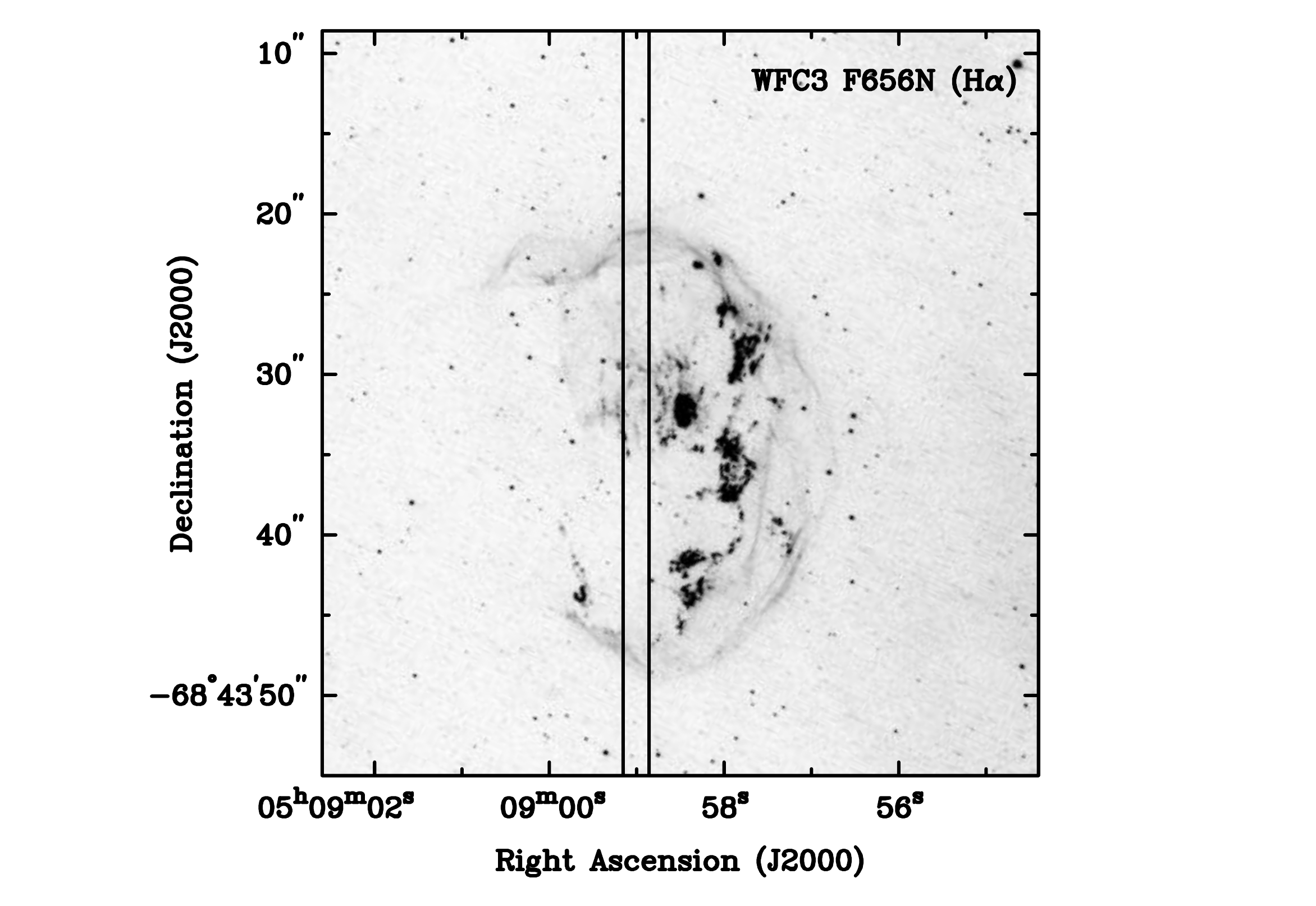}
\epsscale{0.53}
\hspace*{-2cm}\plotone{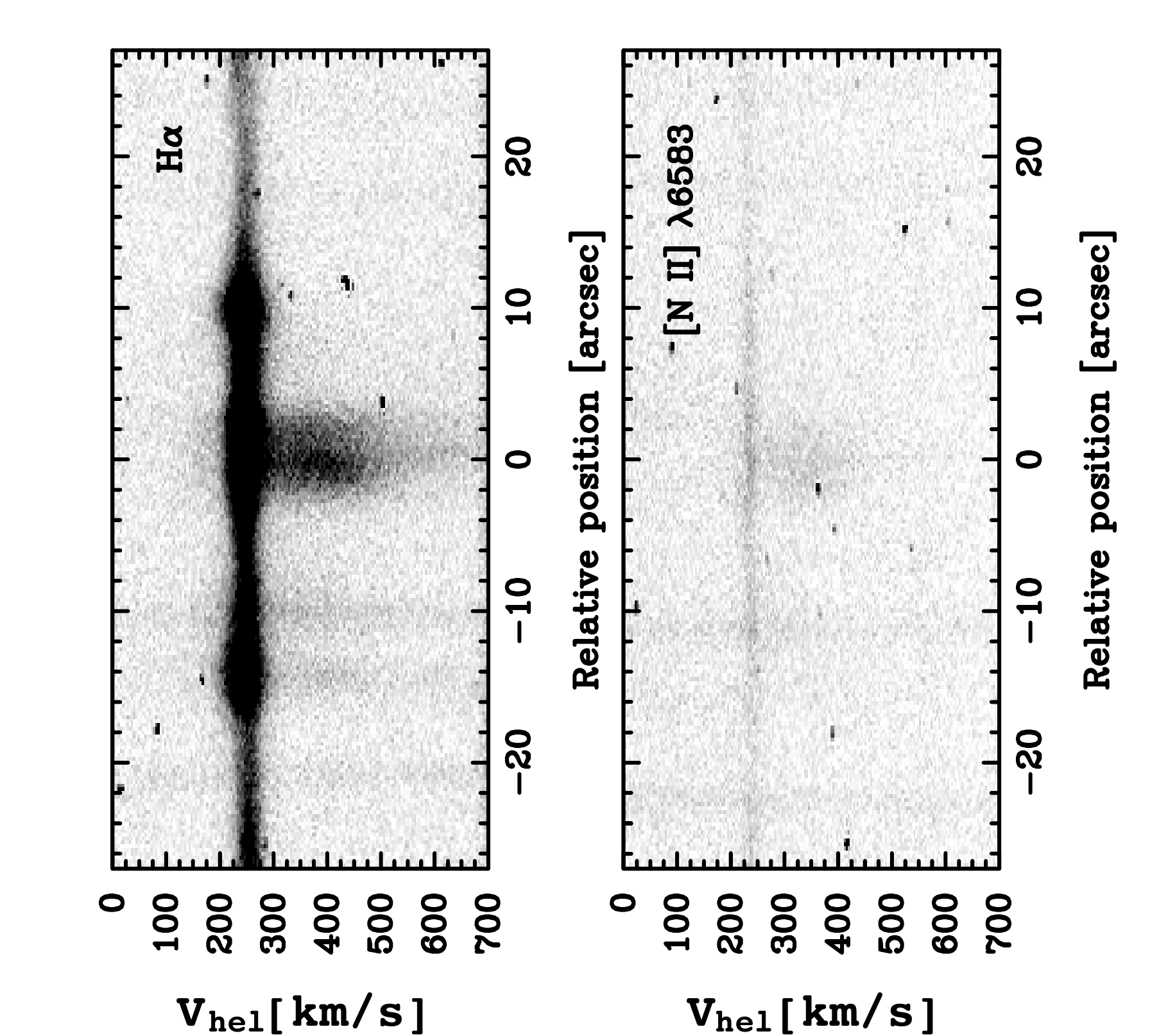}
\caption{Same as Figure \ref{fig:slit_WE}, but at slit position angle 0\degr.}
\label{fig:slit_NS}
\end{figure*}

The long-slit high-dispersion echelle observations provide not only 
kinematic information of ionized gas, but also the [\ion{N}{2}]/H$\alpha$ 
ratio information for the different kinematic components.  We can
compare the long-slit echelle spectral images with the \emph{HST}
H$\alpha$ images to make associations between kinematic features 
in the echelle data and morphological features in the \emph{HST} 
H$\alpha$ images.  Figures \ref{fig:slit_WE} and \ref{fig:slit_NS} show the H$\alpha$ and 
[\ion{N}{2}] $\lambda$6583 line images alongside an \emph{HST} 
H$\alpha$ image overlaid with the corresponding slit positions at
position angles of 44\degr\ and 0\degr, respectively.

Comparisons between the echelle line images and the H$\alpha$ image
clearly suggest three distinct gaseous components:  background 
interstellar gas, Balmer-dominated SNR shell, and shocked nebular
knots in the SNR interior.  A close inspection of echelle line images
along the NS slit (Figure \ref{fig:slit_NS}) shows that the background interstellar 
gas actually shows an incomplete expanding shell structure that appears
to be centered at the SNR.  These four components are detailed below.

\subsection{Background Interstellar Gas}

The echelle line images in Figures \ref{fig:slit_WE} and \ref{fig:slit_NS} show a narrow, nearly
constant-velocity H$\alpha$ component that is present both inside and 
outside the SNR boundaries.  The narrow velocity profile and the spatial
distribution indicate that this ionized gas component has an extended
distribution and is not interacting with the SNR N103B.  The surface 
brightness gradient along the slit (Figure \ref{fig:slit_WE}) is consistent with the 
large-scale H$\alpha$ surface brightness variation shown in Figure \ref{fig:ground_hst},
brighter toward the superbubble of NGC\,1850.

The SNR N103B, however, may not be physically associated with this ionized gas
component, as its Balmer-dominated shell (discussed in the next subsection)
requires the SNR to be in a mostly neutral ISM \citep{chevalier1980}.  
We identify this narrow, 
nearly constant-velocity H$\alpha$ component as the background interstellar gas.
The average [\ion{N}{2}]/H$\alpha$ ratio of this background gas is 0.14 $\pm$ 0.01,
and the average H$\alpha$ velocity profiles have FWHM $\sim$ 30 km s$^{-1}$.

\subsection{Balmer-Dominated SNR Shell}

\begin{figure*}  
\epsscale{0.85}
\plotone{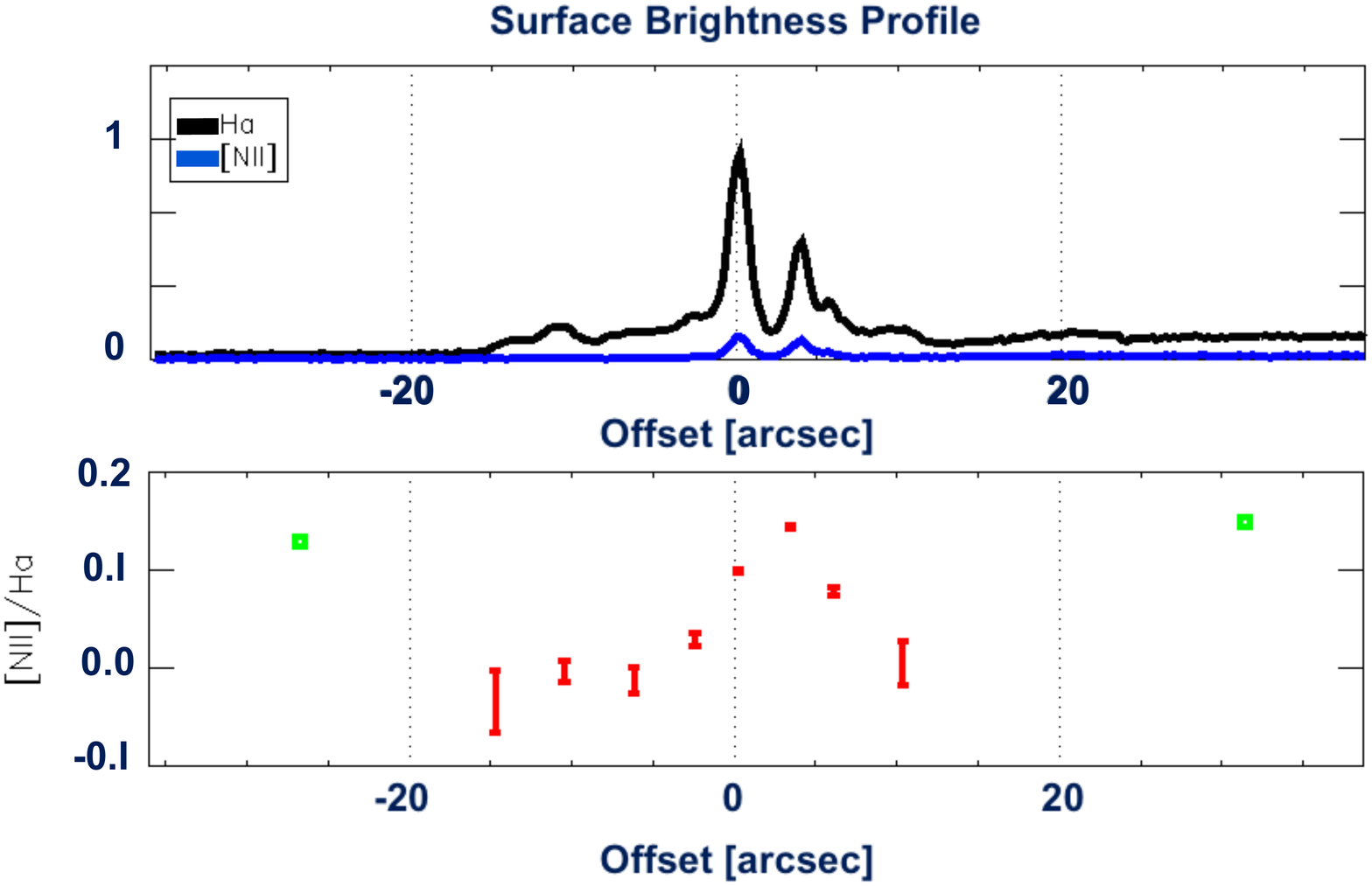}
\caption{(Top) The surface brightness profile of the H$\alpha$ and
[\ion{N}{2}] $\lambda$6583 lines along the echelle slit at 44\degr\
position angle.  The intensity is integrated over 
$\Delta V$ = 100 km s$^{-1}$ centered at $V_{\rm hel} = 250$ km s$^{-1}$.
The two highest peaks correspond to two groups of dense nebular knots.
The two lower peaks at +10$''$ and $-12''$ correspond to the Balmer-dominated
shell rim.
(Bottom) [\ion{N}{2}]/H$\alpha$ ratios along the slit.  The background ISM
component has been subtracted from N103B to derive the ratios
plotted in red.  The [\ion{N}{2}]/H$\alpha$ ratios of the background 
ISM component are plotted in green.}
\label{fig:balmer}
\end{figure*}

Within the SNR boundary, there exists enhanced H$\alpha$ emission at 
velocities similar to that of the background ionized interstellar gas, but
its velocity profiles have broader FWHM, $\sim$ 40 km s$^{-1}$, and 
its surface brightness variations correlate well with that of the filamentary 
elliptical H$\alpha$ shell of N103B.  Unlike the background interstellar gas,
this broad H$\alpha$ shell component does not have [\ion{N}{2}] counterparts
(Figures \ref{fig:slit_WE}, \ref{fig:slit_NS}, \ref{fig:balmer}). This is a characteristic of Balmer-dominated Type Ia SNR 
\citep{tuochy1982}, and has been reproduced by models 
of collisionless SNR shocks into a mostly neutral medium 
\citep{chevalier1980}.
A Balmer-dominated component should exhibit a core and very broad 
wings as shown in low-dispersion spectra \citep{smith1991};
however, our high-dispersion echelle spectra are not sensitive enough 
to detect the broad wings.  Our intermediate-dispersion spectra of N103B
did detect the broad wings, and these will be presented in a future paper.

\begin{figure*}  
\epsscale{0.85}
\plotone{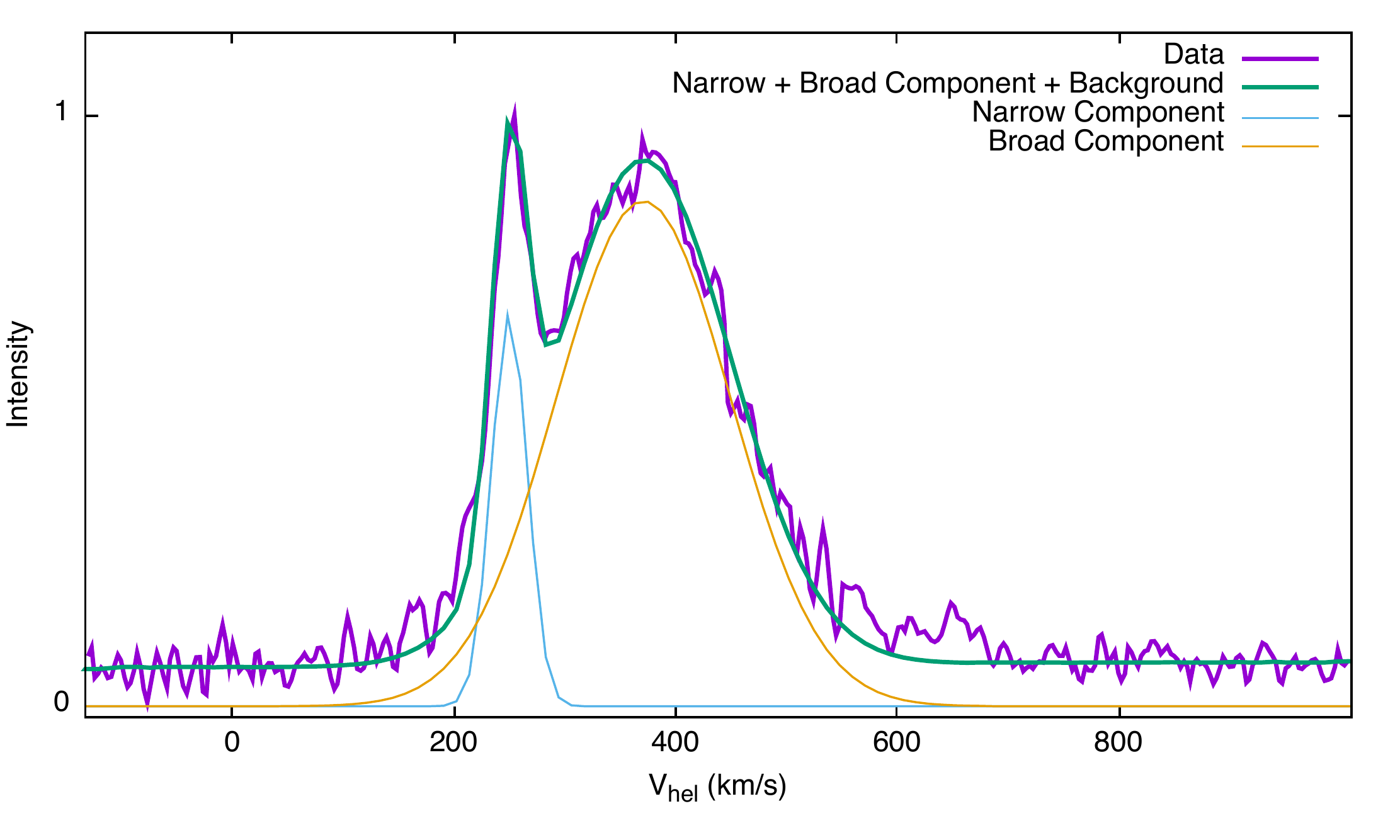}
\caption{CHIRON H$\alpha$ profile of the bright knots in Group I.  The profile is fitted by two Gaussian components: a narrow one from the background ionized gas and a broad one from the shocked CSM.}
\label{fig:chiron_ha}
\end{figure*}

\subsection{Shocked Nebular Knots}

The most spectacular kinematic features in the echelle line images are
discrete regions of H$\alpha$ emission over velocity spreads up to 
$\sim$500 km s$^{-1}$ (Figure \ref{fig:slit_WE}).
These regions correspond to the nebular knots resolved in the \emph{HST}
H$\alpha$ images.  These broad H$\alpha$ components have [\ion{N}{2}]
counterparts, although the velocity spreads of the [\ion{N}{2}] line appear 
smaller due to their lower signal-to-noise ratios.  The [\ion{N}{2}]/H$\alpha$ 
ratios of these broad components are up to 0.18.

It is remarkable that most of the shocked material in the nebular knots
are red-shifted with respect to the background or the Balmer shell component.
The distribution of the CSM is clearly not spherically symmetric; it is one-sided
along the line-of-sight, as well as along the east-west direction.

To illustrate the asymmetric distribution of the CSM, we display in Figure \ref{fig:chiron_ha} the high-dispersion CHIRON spectrum at the H$\alpha$ line of the bright nebular knots in Group I.  This H$\alpha$ line profile can be fitted with two Gaussian components, a narrow one centered near V$_{\rm{hel}} \sim 250$ km s$^{-1}$ with FWHM $\sim$ 40 km s$^{-1}$ and a broad one centered near V$_{\rm{hel}} \sim 370$ km s$^{-1}$ with FWHM $\sim$ 190 km s$^{-1}$.  The two Gaussian components are also plotted in Figure \ref{fig:chiron_ha}.  It can be seen that the broad Gaussian component can describe the bulk of shocked gas but not near the highest velocities, as there is still emission near V$_{\rm{hel}} = 600-700$ km s$^{-1}$.

\begin{figure*} 
\epsscale{1.2}
\hspace*{-1.1cm}\plotone{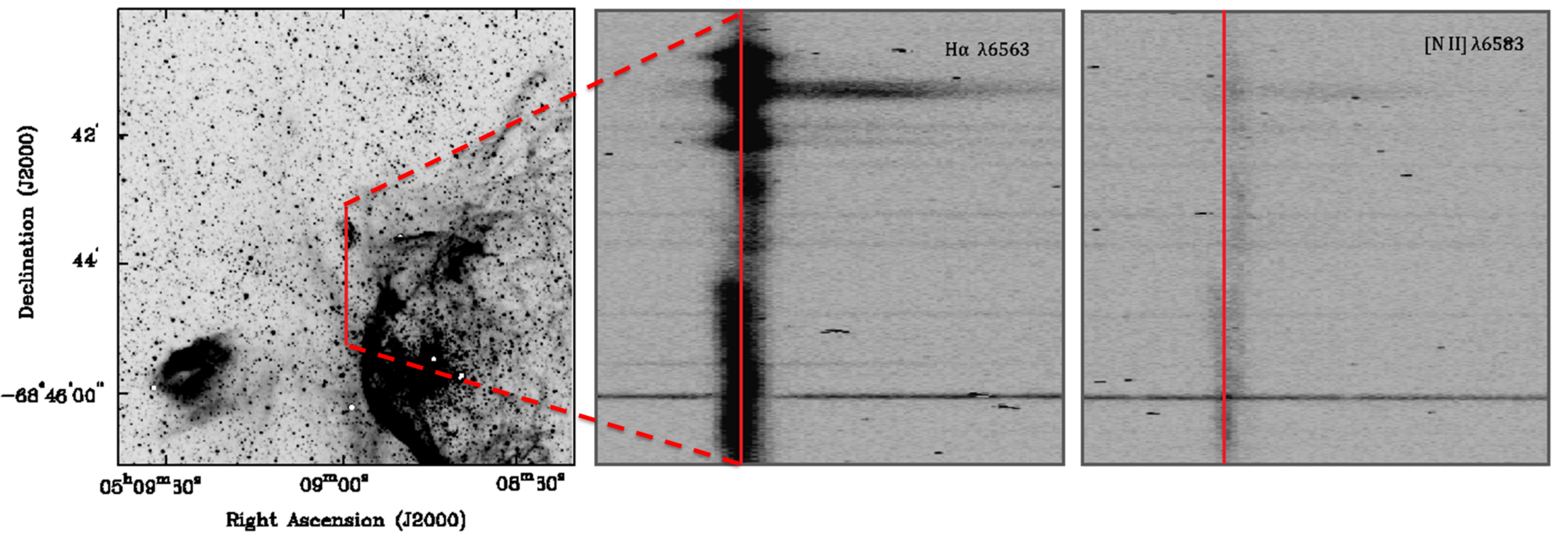}
\caption{H$\alpha$ image (left panel) and long-slot echellograms of the 
H$\alpha$ and[\ion{N}{2}] lines (two right panels) of N103B.  
The N-S slit position is marked in the H$\alpha$ image.
The echellograms have been stretched in the dispersion direction to 
show the velocity structures; a vertical line is drawn at the central
velocity in the echellograms for reference.
The H$\alpha$ line shows the shocks in the N103B SNR well.
The [\ion{N}{2}] line has a smaller thermal width and thus shows well
the line-split resulting from the 10 km s$^{-1}$ expansion in the 
background ISM.}
\label{fig:large_shell}
\end{figure*}

\subsection{Large Incomplete Interstellar Shell}

The [\ion{N}{2}] $\lambda$6583 line, due to nitrogen's larger atomic weight, 
has a smaller thermal width than the H$\alpha$ line, and thus reveals slowly 
expanding structures more clearly.  Figure \ref{fig:large_shell} shows a 
stretched echelle image of the [\ion{N}{2}] line along the NS slit.  It can be seen 
that the small velocity variations in the interstellar component follow a curve 
that is characteristic for an expanding shell.  While the receding side of the shell
is regular and complete, the approaching side is incomplete by more than 50\% of
the spatial extent.  The expansion velocity of this shell is $\sim$10 km s$^{-1}$.
This incomplete shell structure is reminiscent of the incomplete filamentary 
elliptical shell of N103B, and the preferential existence of red-shifted material 
is similar  to what is observed in the shocked nebular knots.  However, a comparison
of its spatial extent and the features in H$\alpha$ image shows that this expanding
shell structure is bounded by two curved filaments extending outwards from the 
superbubble around NGC\,1850.  Therefore, this expanding shell is likely a blister
on the rim of the superbubble, and its large size precludes its physical association 
with the SNR N103B.

\section{Discussion}

\subsection{Structure and Formation of the SNR N103B}

To formulate a feasible model for the structure and formation of the SNR 
N103B, we need to explain the following observed features:  
(1) the apparently elliptical Balmer-dominated shell,
(2) dense knots projected within the Balmer shell interior, and
(3) the eastern extension of diffuse X-ray and radio emission that does
not have optical counterparts.

The presence of the Balmer-dominated shell is indicative of collisionless 
shocks in a mostly neutral ISM \citep{chevalier1980}.  
N103B's Balmer shell thus suggests that the SNR is not in the 
ionized ISM surrounding NGC\,1850's superbubble.
However, the Balmer line emission peaks at velocities that are 
similar to those of the background ionized gas, suggesting 
that the medium surrounding N103B and the ionized superbubble
of NGC\,1850 are likely associated with the same interstellar complex.  
 
The Balmer filaments represent the shock fronts and 
their curvatures can be used to assess the SN explosion centre. 
As the projection of a sphere is always a sphere, the apparently 
elliptical shape of the Balmer shell indicates that the shock fronts
have run into a non-uniform medium and are no longer spherically
symmetric.  The elliptical shape can be produced by the projection 
of an elliptical shell or a tilted round torus (or short cylinder).
As the Balmer shell is open to the east, it is unlikely that the Balmer 
shell is a uniform elliptical shell.  Instead, it is more likely that the 
Balmer ``shell'' is actually a torus or a short cylinder with the eastern
section missing.  The projection of a short cylinder appears elliptical 
and the center of the cylinder is projected at the center of the ellipse.  
Therefore, we identify the geometric center of the Balmer shell as the 
site of SN explosion in N103B.

\begin{figure*} 
\epsscale{1}
\hspace*{0cm}
\plottwo{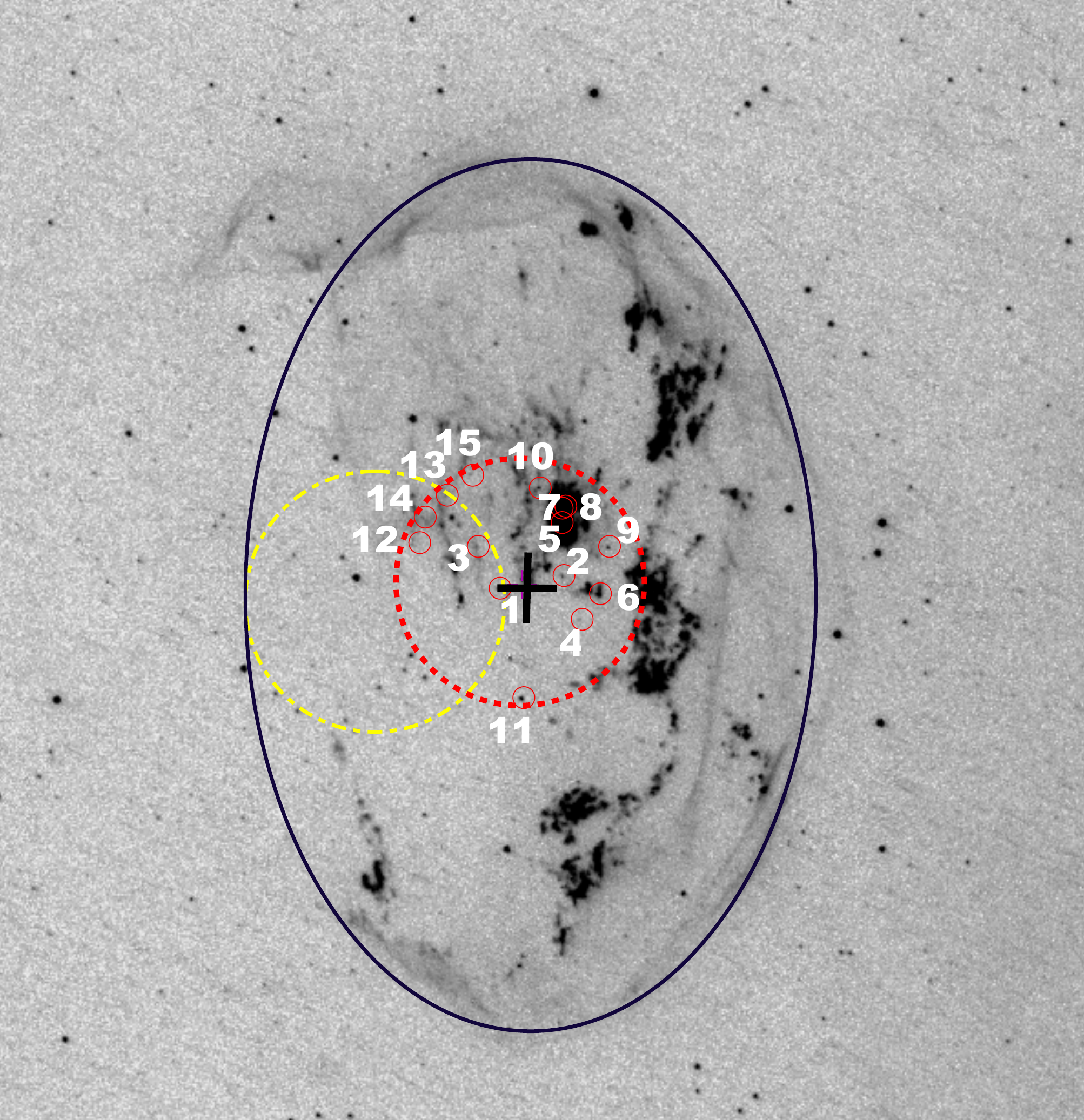}{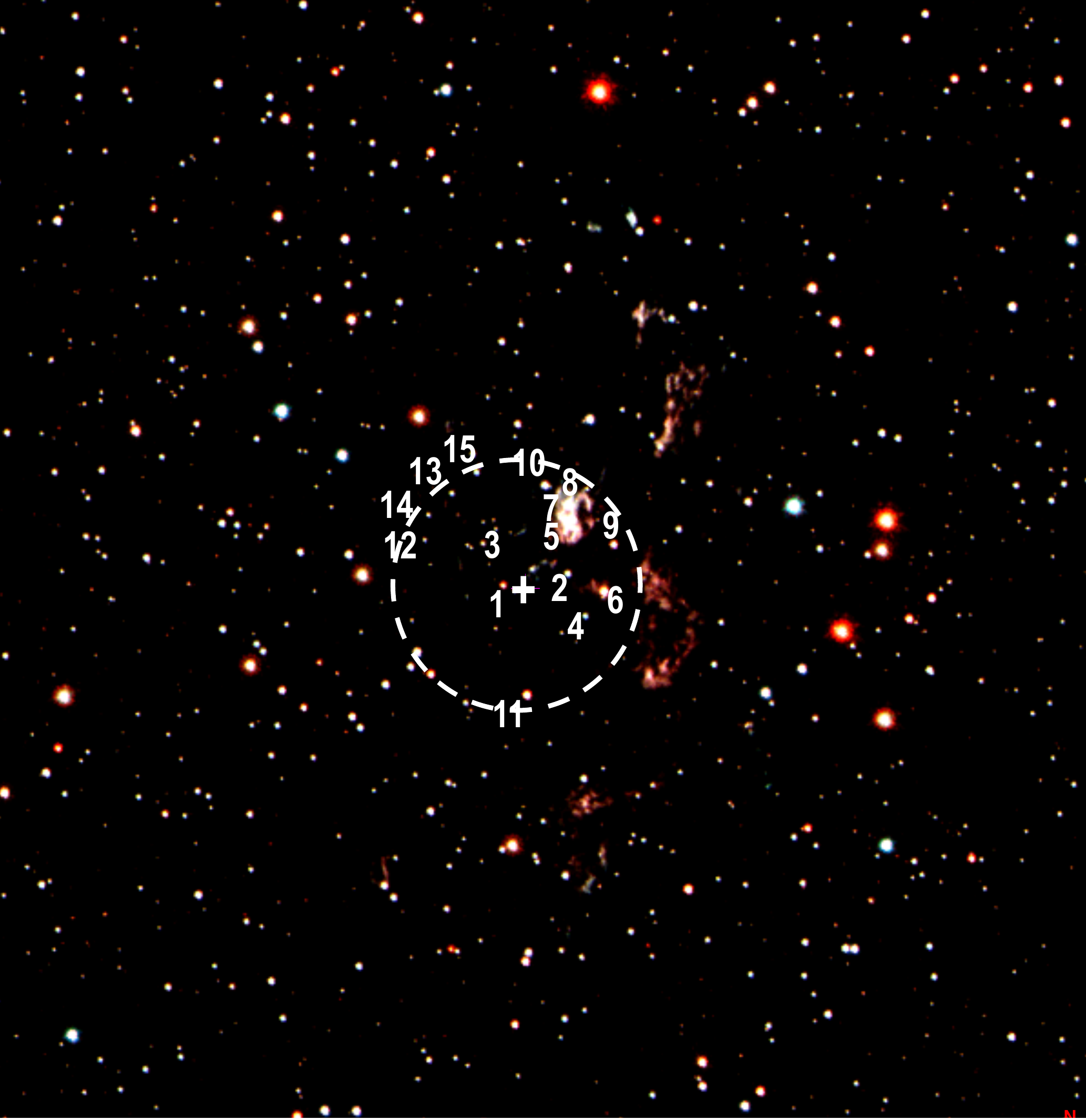}
\caption{(Left) H$\alpha$ image of N103B.  The Balmer-dominated shell is visually 
fitted with a $\sim$28\arcsec\ $\times$ 18\arcsec\ ellipse centered at 05$^{\text{h}}$08$^{\text{m}}$58$^{\text{s}}$.8, 
$-$68$^\circ$43$'$34\farcs7 (J2000).  (Right) Color composite image of N103B
with the F475W image in blue, F555W image in green, and F814W image in red.
The stars with $V < 22.7$ within 4$''$ from the center are marked and numbered 
in both the H$\alpha$ image and the color composite image.  The dashed red circle 
over the H$\alpha$ image illustrates our 4$''$ search radius; the dashed yellow
circle marks the search area of \citet{pagnotta2015}.
}
\label{fig:center}
\end{figure*}

\begin{figure*} 
\epsscale{1}
\hspace*{0cm}
\plotone{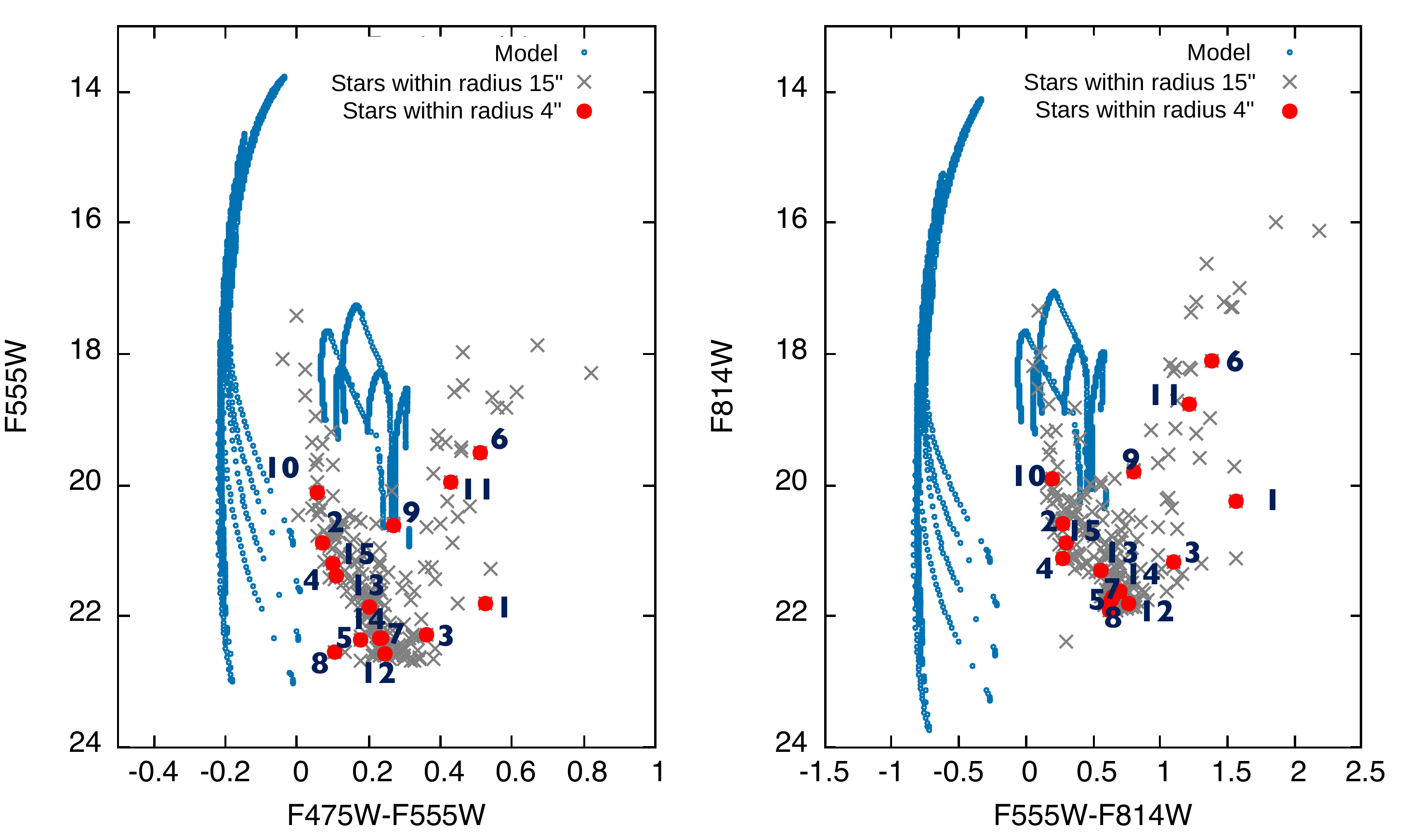}
\caption{Two color-magnitude diagrams of stars with $V < 22.7$ near the explosion 
center of N103B.   The left panel shows $V$ versus $B-V$ and the right panel $I$ versus
$V-I$.  The stars within 4$''$ from the center are plotted in filled circles and marked with
their numbers from Figure \ref{fig:center} and Table 3.  The stars within 15$''$ from the center are 
plotted in grey crosses to illustrate the MS and RG branch of the general background 
stellar population.   The post-impact evolutionary tracks of surviving He-star and MS
companions \citep{pan2014} are to the left and above the MS, respectively.}
\label{fig:cmd}
\end{figure*}

The dense knots seen in the interior of the SNR shell consist of CSM
\citep{williams2014} that has been impacted by the SNR shocks.
The distribution of the knots shows clearly that the circumstellar material
is not spherically symmetric; furthermore, the CSM appears to be distributed in a 
plane.  This geometry is similar to that expected from the model of 
\citet{hachisu2008} for Type Ia SNe.  In this model, a WD accretes
mass from a main sequence companion, and generates a fast wind that 
evacuates material in the polar directions and strips the surface material
of the companion into a torus in the orbital plane. 
This model naturally explains the distribution of the CSM in N103B.  

The CSM is missing towards the east, and the preferentially 
redshifted velocities of the shocked dense knots indicate a lack of CSM in
the direction towards us as well.  This highly asymmetric distribution of  
CSM can be caused by a large proper motion of a mass-losing 
system through the ISM, such as a fast-moving Asymptotic Giant Branch (AGB) 
star \citep{villaver2012}.  As described above, the Balmer shell opens
to the east with no H$\alpha$ emission, but X-ray and radio emission extends
much further east.  This asymmetric distribution of the mostly neutral ISM can 
also be naturally explained by the proper motion of the progenitor binary system.  
In fact, the radio morphology of N103B is reminiscent of the H$\alpha$ image 
of the planetary nebula Sh 2-188, whose morphology has been explained by the 
nebula's high-velocity motion through the ISM \citep{wareing2007}.

We propose that the progenitor of N103B's Type Ia SN has a normal star
companion, and the binary system has a proper motion through the ISM.
The WD's powerful wind strips the companion's envelope to form a circumstellar
torus that is compressed to higher densities in the direction of proper motion. 
The compressed CSM breaks up and forms knots due to Rayleigh-Taylor 
instabilities, while the CSM in the trailing side has low densities and is 
thus undetected.

\subsection{Search for Surviving Companion}

The surviving companion of N103B's SN progenitor, if exists, should be
at a finite distance from the explosion center, and the distance depends 
on its kick velocity at the explosion.  In a previous surviving companion 
search, \citet{pagnotta2015} adopted the average geometric center
N103B's images in radio \citep{dickel1995} and 
X-ray \citep{lewis2003} wavelengths, although these images show very 
uneven surface brightnesses, with the eastern half much fainter than the 
western half.
They found 8 stars within 4$''$ from this explosion center; however, an
unambiguous identification of the surviving companion still awaits  
spectroscopic data demonstrating anomalous chemical abundance or 
kinematic properties, such as high velocity or fast rotation.

Our analysis of the physical structure of N103B, taking into account
both the CSM and ISM, concludes a different explosion center.
In our study, the Balmer-dominated filamentary shell in the \emph{HST} 
H$\alpha$ image is used to determine the site of SN explosion.
The Balmer-dominated shell is visually fitted by an ellipse with a dimension of 
$\sim$28\arcsec\ $\times$ 18\arcsec\ and a center at
05h08m58.8s, $-$68$^\circ$43$'$34\farcs7 (J2000) (see left panel in Figure \ref{fig:center}). 
The apparent ellipticity can be reproduced by a short round cylinder
inclined by $\sim$50$^\circ$ against the sky plane.

Adopting the center of the Balmer-dominated elliptical shell as the explosion
center, we have compiled all stars with $V <$ 22.7 within 4$''$ from the center.
Adopting the age 860 yr derived from the SN light echos \citep{rest2005}, 
these limits correspond to a F8V -- G0V star and a kick velocity of 
(1 pc) / (860 yr) $\sim$ 1140 km s$^{-1}$, respectively.
These stars are listed in Table 3 in order of increasing distance from the 
explosion center.  

To identify candidates for surviving companion of N103B's SN progenitor,
we have plotted in Figure \ref{fig:cmd} all stars with $V < 22.7$ within 15$''$ from 
the explosion center in the $V$ versus $B-V$ and $I$ versus $V-I$ 
CMDs, with the stars within 4$''$ from the 
center marked and numbered in the CMDs as well as in the H$\alpha$ 
and color composite images in Figure \ref{fig:center}.  We have also overplotted the 
post-impact evolutionary tracks of surviving companions for cases of He stars and 
main sequence stars \citep{pan2014}.  

It can be seen that within 
4$''$ from the explosion center, no stars match the expectation of models 
of \citet{pan2014}.  Only star 9 is close to the base of the tracks for 
MS companions in the $V$ versus $B-V$ CMD, but not in the 
$I$ versus $V-I$ CMD; furthermore, star 9 is 3$''$ from the explosion center
and it would need a very high kick velocity, 850 km s$^{-1}$, which is unlikely.

In the SD scenario for Type Ia SNe, the companions may be MS stars, 
RGs, or He stars.  \citet{pan2014} modelled post-impact 
evolution of MS and He companions, but not RG companions.
\citet{podsiadlowski2003}, on the other hand, considered a 1 $M_\odot$ 
subgiant companion which had been stripped of 0.2--0.8 $M_\odot$ envelope
mass and heated by the impact of SN ejecta, and found that the surviving 
companion would be a $\sim$5,000 K subgiant with 1--100 $L_\odot$.  
Interestingly, the star closest to the explosion 
center, star 1, appears to have temperature and luminosity consistent with 
this prediction. Assuming negligible extinction, star 1's $B-V \sim$ 0.53 
suggests a spectral type of late F to early G type, and its $V-I \sim$ 1.56 
suggests a spectral type of late K type;
in the $V$ band, star 1 is 4 times brighter than the Sun, while in the $I$ band
star 1 is 9 times as bright as the Sun.  The mass of the CSM in N103B, as
determined in Section 4, is $\sim$1 $M_\odot$, of similar order of magnitude 
as that of stripped envelope of a subgiant or a giant.  
Furthermore, as shown in Figure \ref{fig:cmd}, 
the location of star 1 in the CMDs is not well populated by LMC or Galactic 
stars, implying a less likelihood to be a popular background star.  Thus, we 
suggest that star 1 is the most likely candidate for the surviving 
companion of the SN progenitor of N103B.   Physical properties of post-impact
companion stars have been modelled, as shown in Table 4 
\citep{pan2013}.
Spectroscopic observations of star 1 in N103B should be made to search 
for high radial velocity and/or fast rotation to confirm or reject its identification
as a surviving companion of the SN progenitor.

\begin{deluxetable*}{ccccccccc}
\tablecolumns{2}
\tabletypesize{\scriptsize}
\tablewidth{0pc}
\tablecaption{Stars Brighter Than $V = 22.7$ and within 4$''$ from the Explosion Center in the SNR N103B}
\tablehead{Star & RA (J2000)   & Dec. (J2000) & B  & V & I & B-V & V-I & $\Theta$}
\startdata
1  & 05:08:58.95 & -68:43:34.7 & 22.33 $\pm$ 0.01 & 21.81 $\pm$ 0.01 & 20.24 $\pm$ 0.01 & 0.53 $\pm$ 0.01 & 1.56 $\pm$ 0.01 & 
0\farcs8\\
2 & 05:08:58.56 & -68:43:34.3 & 20.95 $\pm$ 0.00 & 20.87  $\pm$ 0.00 & 20.59 $\pm$ 0.01 & 0.07 $\pm$ 0.01 & 0.28 $\pm$ 0.01 & 
1\farcs3\\
3 & 05:08:59.06 &  -68:43:33.3 & 22.64 $\pm$ 0.01& 22.28 $\pm$ 0.01 & 21.18 $\pm$ 0.01 & 0.36 $\pm$ 0.02 & 1.10 $\pm$ 0.01 & 
2\farcs0\\
4 & 05:08:58.46 & -68:43:35.8 & 21.51 $\pm$ 0.01& 21.40  $\pm$ 0.01 & 21.12 $\pm$ 0.01 & 0.11 $\pm$ 0.01 & 0.28 $\pm$ 0.01 & 
2\farcs2\\
5 & 05:08:58.57  & -68:43:32.6 & 22.55 $\pm$ 0.01 & 22.38  $\pm$ 0.01 & 21.75 $\pm$ 0.02 & 0.18 $\pm$ 0.02 & 0.63 $\pm$ 0.07 & 
2\farcs4\\
6 & 05:08:58.33 & -68:43:35.0 & 20.01 $\pm$ 0.00& 19.50 $\pm$ 0.00 & 18.12 $\pm$ 0.00 &  0.51 $\pm$ 0.00 & 1.38 $\pm$ 0.00 & 
2\farcs5\\
7 & 05:08:58.56  & -68:43:32.2 & 22.58 $\pm$ 0.01 & 22.34 $\pm$ 0.01 & 21.70 $\pm$ 0.02 & 0.24 $\pm$ 0.02 & 0.64 $\pm$ 0.02 & 
2\farcs9\\
8 & 05:08:58.54  & -68:43:32.1 & 22.66 $\pm$ 0.01& 22.56  $\pm$ 0.01 & 21.97 $\pm$ 0.02& 0.11 $\pm$ 0.02 & 0.63 $\pm$ 0.02 & 
2\farcs9\\
9 & 05:08:58.27 & -68:43:33.4 & 20.89 $\pm$ 0.00 & 20.61  $\pm$ 0.00 & 19.81 $\pm$ 0.00 & 0.27 $\pm$ 0.01 & 0.80 $\pm$ 0.00 & 
3\farcs0\\
10 & 05:08:58.69  & -68:43:31.6 & 20.16 $\pm$ 0.00 & 20.11 $\pm$ 0.00 & 19.91 $\pm$ 0.00 & 0.05 $\pm$ 0.00 & 0.20 $\pm$ 0.01 & 
3\farcs2\\
11 & 05:08:58.80 &  -68:43:38.4 & 20.40 $\pm$ 0.00 & 19.97 $\pm$ 0.00 & 18.76 $\pm$ 0.00 & 0.43 $\pm$ 0.00 & 1.21 $\pm$ 0.00 
&3\farcs6\\
12 & 05:08:59.40 & -68:43:33.3 & 22.82 $\pm$ 0.01 & 22.58 $\pm$ 0.01 & 21.82 $\pm$ 0.01 & 0.24 $\pm$ 0.01 & 0.76 $\pm$ 0.02 & 
3\farcs6\\
13 &  05:08:59.24  & -68:43:31.8 & 22.07 $\pm$ 0.01 & 21.87 $\pm$ 0.01 & 21.31 $\pm$ 0.01 & 0.20 $\pm$ 0.01 & 0.55 $\pm$ 0.01 
&3\farcs8\\
14 & 05:08:59.38  & -68:43:32.5 & 22.56 $\pm$ 0.01 & 22.33 $\pm$ 0.01 & 21.63 $\pm$ 0.01 & 0.23 $\pm$ 0.01 & 0.71 $\pm$ 0.01 & 
3\farcs9\\
15 & 05:08:59.10 & -68:43:31.1 & 21.29 $\pm$ 0.01 & 21.19 $\pm$ 0.01 & 20.89 $\pm$ 0.01 &  0.10 $\pm$ 0.01 & 0.30 $\pm$ 0.01 & 
4\farcs0
\enddata
\end{deluxetable*}

\begin{deluxetable*}{lccc}
\tablecolumns{2}
\tabletypesize{\scriptsize}
\tablewidth{0pc}
\tablecaption{Properties of Post-Impact Companion Stars}
\tablehead{
 & He-Star   & MS-Star   & Giant-Star  
 }
\startdata
Linear velocity (km s$^{-1}$)& 500 - 1000 &  100 - 200 & \textless\, 100  \\
Rotational speed (km s$^{-1}$)& \textgreater\, 50 - 200 & \textless\, 100 & very low  \\
Color & Very blue &   &   \\
Effective Temperature (K)& 10,000 - 100,000 & 5,000 - 10,000  &  ?\\
Stripped Mass ({\it M$_\sun$}) & \textless\, 0.1 & 0.1 - 0.4 &  0.5 - 1.0 
\enddata
\tablerefs{\citet{pan2013}\\ \\\\
}
\end{deluxetable*}

\subsection{Comparison with Kepler's SNR}  \label{sec:comparison}

Kepler's SNR, in the Milky Way at a distance $\sim$5 kpc, resulted from the
SN discovered by Johannes Kepler in 1604.
Optical images of Kepler's SNR display a complex network of knots and filaments
within a radius of $\sim$100\arcsec.  Although the type of Kepler's SN was debated
at first, X-ray, optical, and IR observations of the SNR have pointed toward a 
Type Ia origin \citep{blair2007, reynolds2007}, and demonstrated
the existence of a dense CSM from the progenitor system \citep{vandenbergh1973, 
vandenbergh1977, dennefeld1982, blair1991}. 

N103B and Kepler's SNR share many similar properties.  In X-rays, both SNRs
show prominent Si, S, Ar, Ca, and Fe K$\alpha$ emission and a notable absence of 
O in the spectra of their SN ejecta, as expected for Type Ia SNe 
\citep{hughes1995,reynolds2007}.
In nebular lines at optical wavelengths, both SNRs show similar knots-and-filaments
morphological characteristics: both have Balmer-dominated filaments delineating the outer
boundary of the SNR but only along 1/3 to 1/2 of the periphery, and both exhibit dense
knots distributed within the SNR shell as well as close to the Balmer filaments 
\citep{sankrit2008}.
The dense knots in Kepler's SNR have electron densities of 2,000--10,000 cm$^{-3}$
\citep{blair1991}, and those in N103B $\sim$5,300 cm$^{-3}$. 
The dense knots in Kepler's SNR have been suggested to be CSM ejected by
the SN progenitor system because of the enhanced N abundance implied by
the high [\ion{N}{2}]/H$\alpha$ ratio, $\sim$2 \citep{dennefeld1982, 
dodorico1986}.   The high densities of N103B's knots have been used to argue 
for a CSM origin, similar to Kepler's knots; however, the [\ion{N}{2}]/H$\alpha$ ratios 
of the dense knots in N103B are only $\sim$0.2, not as high as those of Kepler's 
SNR.  This factor of 10 difference in [\ion{N}{2}]/H$\alpha$ ratio cannot be explained
completely by the low N abundance in the LMC, 0.16 times that of the Sun or 0.37 
times that of the solar neighborhood \citep{russell1992}.  The CSM originates
from the envelope of the non-degenerate companion in the SN progenitor system
\citep{hachisu2008}.  
As the CNO process can build up N abundance and the
processed material can be mixed into the envelope through convection,
it is possible that the non-degenerate companion of Kepler's SN progenitor has a higher
mass than that for N103B, and Kepler's CSM contains CNO-processed material.

A search for surviving companion of Kepler's SN progenitor has been conducted by
\citet{kerzendorf2014} using spectroscopic observations of the 24 brightest stars
near the center of  Kepler's SNR.  From the radial velocities of these stars, they ruled 
out a RG companion with high certainty, and found no candidates for a donor star
with luminosity $>$10 $L_\odot$, although follow-up astrometric observations for
proper motion and high-dispersion spectra for stellar rotation are needed for further 
search.  In this paper, we identify a possible candidate for surviving companion of
N103B's SN progenitor near the explosion center of the SNR.  This candidate was likely
a solar-mass sub-giant that had no significant CNO cycle burning of H in its interior
\citep{podsiadlowski2003}.
It also needs follow-up spectroscopic observations of radial velocity, rotational
velocity, and abundances to verify its association with N103B's SN progenitor.

\section{Summary} 

N103B is a Type Ia SNR in the LMC.  We have obtained \emph{HST} images of 
N103B in the H$\alpha$ and $BVI$ bands in order to study the physical structure 
of the SNR and to search for candidates of surviving companion of the SN 
progenitor.  We have also obtained CTIO 4m high-dispersion, long-slit echelle 
and 1.5m high-dispersion, multi-order echelle observations of the SNR in 
order to kinematically differentiate between interstellar and circumstellar gas 
components and to detect SNR shocks.

The SNR N103B is projected against the diffuse ionized interstellar gas in
the outskirts of the superbubble around the NGC\,1985 cluster.  The \emph{HST}
H$\alpha$ image shows delicate filaments that form an incomplete $28''\times18''$
elliptical ring, and groups of dense knots interior to the ring.  Both the filaments 
and knots are distributed preferentially on the west side, and exhibit an opening 
to the east where X-ray and radio emission extends further by $\sim15''$.

The H$\alpha$ and [\ion{N}{2}] lines from the echelle data show that the filaments
are dominated by Balmer lines (no forbidden lines) characteristic of collisionless
shocks advancing into a mostly neutral ISM, while the knots have electron densities as
high as 5300 cm$^{-3}$ and are shocked CSM.  We suggest that the 
asymmetric distribution of the CSM is caused by a large proper motion of the 
progenitor binary system toward the west.  The mass loss from the progenitor 
system is compressed by the ISM to higher densities on the leading side, and 
is dispersed and diluted by the void in the trailing side.

As the Balmer-dominated filaments delineate the SNR shock front into the ISM,
they can be used to trace the location of SN explosion.  The Balmer-dominated
filaments are distributed roughly along an ellipse, and we adopt its center as the
site of SN explosion.  We have constructed CMDs and compare the locations of
stars projected within 4$''$ (1 pc) from the explosion center with the post-shock 
evolutionary tracks of surviving companions calculated by 
\citet{pan2012,pan2013,pan2014}.
No stars projected near the explosion center are coincident with these tracks.
On the other hand, the star closest to the explosion center appears to have
the colors and luminosity predicted by \citet{podsiadlowski2003} for a
1 $M_\odot$ subgiant companion with 0.2--0.8 $M_\odot$ of envelope stripped
and heated by the SN shock.  Future spectroscopic observations of radial velocity
and rotational velocity of this star are needed to confirm or reject its identification 
as N103B SN progenitor's surviving companion.

\acknowledgments
YHC thanks Drs.\ W.\ Kerzendorf, K.\ Nomoto, and P.\ Podsiadlowski for 
enlightening discussions during the Supernovae Workshop at the Munich 
Institute for Astro- and Particle Physics (MIAPP) of the DFG cluster of excellence
``Origin and Structure of the Universe.''  YHC acknowledges the support and
hospitality of MIAPP.  We also thank Dr. Ronald Webbink for reading and
commenting on this paper.
Access to the 1.5m/Chiron, operated by the SMARTS Consortium, was made possible
by support from the Provost of Stony Brook University, Dennis Assanis.
This project is supported by the NASA grant HST-GO-13282.01-A.
YHC and CJL are supported by Taiwanese Ministry of Science and 
Technology grant MOST 104-2112-M-001-044-MY3.

\end{document}